\documentclass{wscpaperproc}   
\usepackage{latexsym}
\usepackage{graphicx}
\usepackage{mathptmx}
\usepackage{color}

\usepackage{comment}
\usepackage{graphicx}
\usepackage{xargs}
\usepackage{ifthen}
\usepackage{algorithm}
\usepackage{algpseudocode}
\def\argmax{\mathrm{argmax}}
\def\argmin{\mathrm{argmin}}

\usepackage{subfig,pgfplots}

\pgfplotsset{axisStyle/.style={
	enlargelimits=false,  ylabel near ticks, xlabel near ticks, axis line style={draw=none},  tick style={draw=none}, yticklabel={\empty}, xticklabel={\empty}, ylabel style={xshift=0.2cm,}, xlabel style={xshift=0.3cm, yshift=0.2cm},
} } 

\usepackage{amsmath}
\usepackage{amsfonts}
\usepackage{amssymb}
\usepackage{amsbsy}
\usepackage{amsthm}

\usepackage[pdftex,colorlinks=true,urlcolor=blue,citecolor=black,anchorcolor=black,linkcolor=black]{hyperref}

\newtheoremstyle{wsc}
{3pt}
{3pt}
{}
{}
{\bf}
{}
{.5em}
{}

\theoremstyle{wsc}
\newtheorem{theorem}{Theorem}

\def\tcb{\textcolor{black}}

\newcommand{\X}{\mathcal{S}}   

\newcommandx*{\x}[1][1={}]{#1}  

\newcommandx*{\xStar}[1][1={t}]{k({#1})}  

\newcommandx*{\xC}[1][1={t}]{c({#1})}  

\newcommandx*{\xG}[1][1={t}]{g({#1})}  

\newcommandx*{\m}[1][1={}]{\mu_{#1}}  

\newcommandx*{\p}[1][1={}]{\theta_{#1}}  

\newcommandx*{\Pm}[2][1={}, 2={t}]{\bar{\mu}_{#1}({#2})}  

\newcommandx*{\Pp}[2][1={}, 2={t}]{\theta_{#1}({#2})}  

\newcommandx*{\sample}[1][1={t}]{x({#1})}  

\newcommandx*{\obs}[1][1={t+1}]{Y({#1})}  

\newcommandx*{\Em}[2][1={}, 2={t}] {\ifthenelse{\equal{#2}{}}{\bar{Y}_{#1}}{\bar{Y}_{#1}({#2})}}  

\newcommandx*{\Emt}[2][1={}, 2={t}]{\hat{Y}_{#1}({#2})}  

\newcommandx*{\Isa}[1][1={}]{\sigma_{#1}}  

\newcommandx*{\Iva}[1][1={}]{\sigma_{#1}^2}  

\newcommandx*{\Ip}[1][1={}]{\lambda_{#1}}  
 
\newcommandx*{\tnr}[2][1={i}, 2={t}]{r_{#1}({#2})}  

\newcommandx*{\all}[2][1={i}, 2={t}]{\ifthenelse{\equal{#2}{*}}{\alpha_{#1}^{#2}}{\alpha_{#1}({#2})}}  

\newcommandx*{\cnr}[1][1={i}]{r_{#1}}  

\newcommand{\bdgt}{R}

\newcommandx*{\ei}[2][1={}, 2={t}]{\text{EI}_{#1}({ #2})}  

\newcommandx*{\cei}[2][1={}, 2={t}]{\ifthenelse{\equal{#2}{}}{\text{CEI}_{#1}}{\text{CEI}_{#1}({ #2})}}  

\newcommand{\normal}{\mathrm{N}}  

\newcommandx*{\salg}[1][1={t}]{\mathcal{F}^{#1}}  

\newcommandx*{\ind}[1][1={}]{\mathbb{I}_{\{#1\}}}  

\newcommandx*{\Prob}[1]{\mathbb{P}\left\{ #1\right\}}  

\newcommandx*{\E}[1]{\mathbb{E}\left[#1\right]}  

\newcommandx*{\Cov}[2][1, 2={}]{\mathrm{Cov}\left( #1\right)}  

\newcommandx*{\At}[2][1={}, 2={t}]{A_{#1}({#2})}  

\newcommandx*{\xit}[2][1={}, 2={t}]{\xi_{#1}({#2})}  

\allowdisplaybreaks

\title{gCEI} 
\begin{document}

\pagestyle{fancyplain}

\thispagestyle{plain}
\firstPageHead{}

\chead{\fancyplain{}{\itshape Avci, Nelson, and W\"{a}chter}}

\rhead{}
\cfoot{}
\renewcommand{\headrulewidth}{0pt} 

\makeatletter
\let\@internalcite\cite
\def\cite{\def\@citeseppen{-1000}%
    \def\@cite##1##2{(##1\if@tempswa , ##2\fi)}%
    \def\citeauthoryear##1##2##3{##1 ##3}\@internalcite}
\def\citeNP{\def\@citeseppen{-1000}%
    \def\@cite##1##2{##1\if@tempswa , ##2\fi}%
    \def\citeauthoryear##1##2##3{##1 ##3}\@internalcite}
\def\citeN{\def\@citeseppen{-1000}%
    \def\@cite##1##2{##1\if@tempswa, ##2)\else{}\fi}%
    \def\citeauthoryear##1##2##3{##1 (##3)}\@citedata}
\def\citeA{\def\@citeseppen{-1000}%
    \def\@cite##1##2{(##1\if@tempswa , ##2\fi)}%
    \def\citeauthoryear##1##2##3{##1}\@internalcite}
\def\citeANP{\def\@citeseppen{-1000}%
    \def\@cite##1##2{##1\if@tempswa , ##2\fi}%
    \def\citeauthoryear##1##2##3{##1}\@internalcite}
\def\shortcite{\def\@citeseppen{-1000}%
    \def\@cite##1##2{(##1\if@tempswa , ##2\fi)}%
    \def\citeauthoryear##1##2##3{##2 ##3}\@internalcite}
\def\shortciteNP{\def\@citeseppen{-1000}%
    \def\@cite##1##2{##1\if@tempswa , ##2\fi}%
    \def\citeauthoryear##1##2##3{##2 ##3}\@internalcite}
\def\shortciteN{\def\@citeseppen{-1000}%
    \def\@cite##1##2{##1\if@tempswa, ##2\else{}\fi}%
    \def\citeauthoryear##1##2##3{##2 (##3)}\@citedata}
\def\shortciteA{\def\@citeseppen{-1000}%
    \def\@cite##1##2{(##1\if@tempswa , ##2\fi)}%
    \def\citeauthoryear##1##2##3{##2}\@internalcite}
\def\shortciteANP{\def\@citeseppen{-1000}%
    \def\@cite##1##2{##1\if@tempswa , ##2\fi}%
    \def\citeauthoryear##1##2##3{##2}\@internalcite}
\def\citeyear{\def\@citeseppen{-1000}%
    \def\@cite##1##2{(##1\if@tempswa , ##2\fi)}%
    \def\citeauthoryear##1##2##3{##3}\@citedata}
\def\citeyearNP{\def\@citeseppen{-1000}%
    \def\@cite##1##2{##1\if@tempswa , ##2\fi}%
    \def\citeauthoryear##1##2##3{##3}\@citedata}
%
%
%
\def\@citedata{%
    \@ifnextchar [{\@tempswatrue\@citedatax}%
                  {\@tempswafalse\@citedatax[]}%
}

\def\@citedatax[#1]#2{%
\if@filesw\immediate\write\@auxout{\string\citation{#2}}\fi%
  \def\@citea{}\@cite{\@for\@citeb:=#2\do%
    {\@citea\def\@citea{, }\@ifundefined
       {b@\@citeb}{{\bf ?}%
       \@warning{Citation `\@citeb' on page \thepage \space undefined}}%
{\csname b@\@citeb\endcsname}}}{#1}}%

%
\def\@citex[#1]#2{%
\if@filesw\immediate\write\@auxout{\string\citation{#2}}\fi%
  \def\@citea{}\@cite{\@for\@citeb:=#2\do%
    {\@citea\def\@citea{; }\@ifundefined
       {b@\@citeb}{{\bf ?}%
       \@warning{Citation `\@citeb' on page \thepage \space undefined}}%
{\csname b@\@citeb\endcsname}}}{#1}}%

%
\def\@biblabel#1{}
\makeatother



\newdimen\bibindent
\bibindent=0.0em
\def\thebibliography#1{\section*{\refname}\list
   {}{\settowidth\labelwidth{[#1]}
   \leftmargin\parindent
   \itemindent -\parindent
   \listparindent \itemindent
   \itemsep 0pt
   \parsep 0pt}
   \def\newblock{}
   \sloppy
   \sfcode`\.=1000\relax}


\setlength{\baselineskip}{12.7pt}

\title{GETTING TO ``RATE-OPTIMAL'' IN RANKING \& SELECTION}

\author{
Harun Avci\\
Barry L. Nelson\\
Andreas W\"{a}chter\\[12pt]
Department of Industrial Engineering \& Management Sciences\\
Northwestern University\\
Evanston, IL 60208 USA\\
}

\maketitle

\section*{ABSTRACT} 

In their 2004 seminal paper, Glynn and Juneja formally and precisely
established the rate-optimal, probability-of-incorrect-selection,
replication allocation scheme for selecting the best of $k$ simulated
systems. In the case of independent, normally distributed outputs this
allocation has a simple form that depends in an intuitively appealing
way on the true means and variances. Of course the means and
(typically) variances are unknown, but the rate-optimal allocation
provides a target for implementable, dynamic, data-driven policies to
achieve.  In this paper we compare the empirical behavior of four
related replication-allocation policies: mCEI from Chen and Rzyhov and
our new gCEI policy that both converge to the Glynn and Juneja allocation;
AOMAP from Peng and Fu that converges to the OCBA optimal allocation;
and TTTS from Russo that targets the rate of convergence of the
posterior probability of incorrect selection.  We find that these
policies have distinctly different behavior in some settings.

\section{INTRODUCTION}
\label{sec:Introduction}

Ranking and selection (R\&S) is one of the fundamental methods for
solving stochastic simulation optimization problems.  In the canonical
version of the R\&S problem, the aim is to identify the single best
among a finite number ($k$) of systems, where the performance of each
system can only be estimated using simulation output; here ``best''
means the maximum or minimum expected value of performance.  The ideal
R\&S procedure either (a) allocates a limited simulation budget so as
to maximize the likelihood that the best is identified, or (b)
allocates simulation effort as efficiently as possible until a
prespecified likelihood is obtained.  This paper addresses formulation
(a). 

The R\&S literature contains many policies for version (a) that
sequentially obtain replications from systems and adapt as more and
more output data are obtained. These policies tend to be Bayesian or
Bayesian-inspired, and include versions of optimal computing budget
allocation (OCBA, \shortciteNP{Chen2000}), expected improvement (EI,
\shortciteNP{Jones1998}), knowledge gradient (KG,
\citeNP{frazier2007knowledge}), and multi-armed bandits (MAB,
\citeNP{jamieson2014best}). In this paper, an \emph{allocation} is the
fraction of a fixed budget of replications that is assigned to each
simulated system, while a \emph{policy} is an algorithm for
sequentially, and usually adaptively, allocating individual
replications to systems.

\citeN{Glynn2004} derived an expression for the asymptotically optimal
\emph{static} replication allocation by using large-deviations theory.
They represented the replications allocated to system $i$ as $\alpha_i
\bdgt$, where $\bdgt$ is the total budget of replications, $\alpha_i >
0$, and $\sum_{i=1}^k \alpha_i = 1$.  The policy is ``optimal'' in a
sense that the probability of incorrect selection (PICS) decays
exponentially with the best possible exponent as $R$ increases;
incorrect selection means choosing any of the $k-1$ inferior systems.
Unfortunately, the optimal allocation depends on the underlying output
distributions and their parameters, which are typically unknown, and
naive plug-in strategies tend not to work well.  This leads to the
idea of having \emph{adaptive} policies that aggressively pursue the
best system in small samples but converge to an ``optimal''
allocation, such as that of \citeN{Glynn2004}, in the limit.

Recently, several such policies have been proposed. The empirical
allocation of the \emph{modified complete expected improvement} (mCEI)
policy of \citeN{Chen2019} converges to the rate-optimal allocation of
\citeN{Glynn2004} under certain conditions. The \emph{asymptotically
optimal myopic allocation policy} (AOMAP) of \citeN{Peng2017}
converges to the OCBA limiting allocation. And finally, the
\emph{top-two Thompson sampling} (TTTS) policy of \citeN{Russo2020}
seeks the optimal rate of convergence of the posterior probability of
incorrect selection to $0$; TTTS is one of three ``top-two'' policies
identified by \citeN{Russo2020}. \tcb{Notice that AOMAP and TTTS do
\emph{not} converge to the \citeN{Glynn2004} optimal allocation but
instead to limits that are arguably desirable.}

To this list we add our new \emph{gradient of CEI} (gCEI) policy that
attains the same limit as mCEI.  We then empirically evaluate the
fixed-budget behavior of all four policies under assumptions that
support all four: independent, normally distributed output with known
variances.  Fixed-budget, as opposed to asymptotic behavior is what an
analyst actually experiences in practice. It is worth stating that
mCEI, gCEI, AOMAP and TTTS can all be ``beaten'' in some sense by
policies specifically designed for good finite-sample performance,
especially when the number of systems $k$ is very large; \tcb{such
policies fully eliminate apparently inferior systems quickly, while
mCEI, gCEI, AOMAP and TTTS keep all systems in play until the budget
is consumed.} Nevertheless, they are building blocks for more
sophisticated policies so the comparison is relevant.

The remainder of the paper is organized as follows: We provide a brief
literature review in Section~\ref{sec:Background} and formulate the
R\&S problem in Section~\ref{Preliminaries}.  We state the AOMAP, mCEI
and TTTS policies in Section~\ref{Methods}, then introduce gCEI and
sketch a proof of its convergence in Section~\ref{gCEI Algorithm and
Main Results}.  Empirical performance of all four policies is given in
Section~\ref{Numerical Experiments}.  Finally,
Section~\ref{Conclusion} concludes the paper.

\section{ESSENTIAL LITERATURE}
\label{sec:Background}

The EI criterion was first introduced by \shortciteN{Jones1998} for
Bayesian optimization of deterministic simulations.  Adapting EI to
the R\&S problem with independent normal observations,
\citeN{Ryzhov2016} derived the asymptotic sampling allocation implied
by EI and showed that it is related to the OCBA allocation of
\shortciteN{Chen2000}.  Since EI does not achieve an exponential
convergence rate, \citeN{Peng2017} proposed a variant of EI, called
AOMAP.  In the known-variance case, AOMAP achieves the OCBA allocation
of \shortciteN{Chen2000} in the limit.  \citeN{Peng2017} note that an
adjustment can be made to EI to achieve \emph{any} well-defined
limiting allocation, including that of \citeN{Glynn2004}, but this
requires solving for the limiting allocation on each iteration based
on plug-in estimates.


Since EI was originally created for deterministic simulations, it does
not directly account for the uncertainty in the output from a
stochastic simulation.  To incorporate this uncertainty,
\shortciteN{Salemi2019} proposed complete expected improvement (CEI).
\tcb{For the R\&S problem with independent normal observations,
\citeN{Chen2019} presented a modified CEI policy. Similarly, under
more general sampling distributions, \citeN{chen2019balancing}
proposed the balancing optimal large deviations policy that evaluates
the approximate individual large-deviation rate functions and balances
them iteratively.  Both policies asymptotically achieve the optimal
allocations of Glynn and Juneja (2004) when the variances are known,
or when variances are unknown but continually updated via plug-in
estimators.}


More recently, \citeN{Russo2020} proposed three different Bayesian
policies for adaptively allocating measurement effort in stochastic
decision problems including simulation.  On every iteration, these
policies use the posterior distribution of the output parameter (e.g.,
mean) to identify the top-two alternatives; one of them is randomly
chosen to measure (simulate). The selection probability is a tuning
parameter, although \citeN{Russo2020} found $1/2$ had robust empirical
performance.  Top-two probability sampling identifies the two
alternatives with the largest posterior probabilities of being
optimal.  Similarly, top-two value sampling considers the posterior
expected value of the difference between the mean of each system and
the best of the others.  The third version is TTTS; see
\citeN{Thompson1933} for the origins of Thompson sampling. We
employ TTTS with selection probabililty $1/2$ in this paper, and
describe it fully below. For the known variances case,
\citeN{Russo2020} showed that these policies attain the best
exponential rate of convergence of the posterior probability of
incorrect selection for the true best system when the tuning parameter is set
optimally or adjusted adaptively toward the optimal value.

In addition to AOMAP, mCEI and TTTS, we propose a new policy called
gCEI that makes replication-allocation decisions based on the gradient
of CEI with respect to the number of replications obtained from each
system, treating ``number of replications'' as if it were
continuous-valued.  Like mCEI, it achieves the optimal allocation of
\citeN{Glynn2004}.  However, gCEI attains this limit without the need
to directly enforce the balance between simulating the best system and
the inferior systems, as mCEI does.



\section{PRELIMINARIES}
\label{Preliminaries}

Let $\X = \{\x[1], \x[2], \ldots, \x[k]\}$ be the set of systems.
Each system $\x[i] \in \X$ has an unknown mean $\m[i]$.  Bigger
is better, and unknown to us $\m[1] \le \m[2] \le \ldots \le \m[k-1] <
\m[k]$.  Our goal is to find system $\x[k]$ which is the unique
best.

From a Bayesian perspective, the mean of each system $\x[i]$ has a
prior distribution $\m[i] \sim \normal(\Pm[i][0], 1/ \Pp[i][0])$ where
$\Pm[i][0]$ and $\Pp[i][0]$ are the prior mean and precision,
respectively.  The prior mean $\Pm[i][0]$ represents the initial
belief about the true value of $\m[i]$ whereas the prior precision
$\Pp[i][0]$ quantifies the confidence in this belief.  We assume that
$\m[i]$'s are independent of each other under this prior. Notice that
we use $\mu_i$ to denote the true, fixed means of the systems, and
$\Pm[i]$ to denote the posterior mean through iteration $t$,
which is a random variable.


We consider a finite horizon problem with a fixed simulation budget:
Let $\bdgt$ be the length of our finite horizon.  At each iteration $t
=0,1,\ldots, \bdgt$, we obtain a single replication $\obs$ by
simulating $\sample$, an independent and identically distributed
$\normal(\m[{\sample}], \Iva[{\sample}])$ random variable with
$\Iva[i]>0$ being the variance inherent to the stochastic simulation
output for system $\x[i]$.  In this paper we assume that
$\Iva[i]$'s are known and that each system is simulated
independently of the others (no common random numbers).  


Let $\salg$ be the sigma-algebra generated by $\{\sample[\tau],
\obs[\tau+1]\}_{\tau=0}^{t-1} $.  Using the
recursive approach in \citeN{DeGroot1970}, the posterior parameters
for systems $i \in \X$ at iteration $t$ are
\begin{align*}
\Pm[i][t+1]& = \begin{cases} 
\dfrac{ \Pm[i][t] \Pp[i][t] + \obs /\Iva[i]} {\Pp[i][t] + 1 /
\Iva[i]} & \text{ if }  \sample = \x[i] \text{ (i.e., if system $i$ is simulated at iteration $t$) }\\
\Pm[i][t]& \text{ if } \sample \neq \x[i],
\end{cases} \\
\Pp[i][t+1] &= \begin{cases} 
\Pp[i][t] + 1 / \Iva[i] & \text{ if } \sample = \x[i]  \\
\Pp[i][t]& \text{ if } \sample \neq \x[i].
\end{cases}
\end{align*}
Let $\tnr[i]$ denote the total number of replications that have been
obtained by simulating system $\x[i]$ up to iteration $t$, i.e.,
$\tnr[i] = \sum_{\tau=0}^{t-1} \ind [{\sample[\tau] = \x[i]}]$ where
$\ind[\cdot]$ is the indicator function.  We employ a non-informative
prior (i.e., $\Pp[i][0] = 0$).  Thus, we have
$\Pm[i] = \Em[i] $ and $\Pp[i] = \tnr[i] / \Iva[i] $ where
\[
\Em[i] = \frac{1}{\tnr[i]} \sum_{\tau=0}^{t-1}  \ind [{\sample[\tau]
= \x[i]}] \obs[\tau+1] 
\]
is the sample mean of system $\x[i]$.

Let $\x[\xStar]$ be the sample-best system at iteration $t$,
$\x[\xStar] = \argmax_{i \in \X} \{\Pm[i]\}
=\argmax_{i \in \X} \left\{\Em[i] \right\}$.
We define the (frequentist) probability of correct selection (PCS) at
iteration $t$ as $\Prob{ \xStar = k }$; thus, the probability of
incorrect selection (PICS) is $\Prob{ \xStar \neq k }$. These
quantities are with respect to the fixed, true means. We can also
define corresponding quantities for the posterior probability that
system $i$ is or is not the best, which is relevant for TTTS.

A generic adaptive policy is given in
Algorithm~\ref{alg:adaptiveAllocation}.  AOMAP, mCEI, gCEI and TTTS
differ in how they decide $\sample$ in Step~3 to obtain good
finite-$\bdgt$ and asymptotic $\bdgt \rightarrow \infty$ performance.

\begin{algorithm}[h!]
  \caption{Generic Adaptive Policy} \label{alg:adaptiveAllocation}
  \begin{algorithmic}[1]
  \State Let $\sample[0] = \x[i]$ for some $\x[i]\in\X$.
  Obtain $\obs[1]$ and update $\salg[1]$.
  Also, let $t \gets 1$.
  \While{$ t < \bdgt$} 
  \State Decide to simulate $\sample$.
  \State  Obtain $\obs$ by simulating $\sample$, 
update $\salg[t+1] \gets \salg[t] \cup\{\sample, \obs\}$ and $t \gets t + 1$.
  \EndWhile
  \State Return $\x[\xStar][\bdgt] = \argmax_{i \in \X}
  \{\Pm[i][\bdgt]\}$ as the  selected best system.
  \end{algorithmic}
\end{algorithm}

\section{POLICIES}
\label{Methods}

In this section we summarize three different policies in the recent
literature.

\subsection{AOMAP}

For our R\&S problem, the EI for system $\x[i]$ at iteration $t$ is 
\begin{align*}
\ei[{\x[i]}] & = \E{\max\{ \m[i] -  \Pm[\xStar],0\} | \salg}
=\sqrt{ 1 / \Pp[i] } \  f\left(\frac{ \Pm[i] - \Pm[\xStar]}{\sqrt{ 1 / \Pp[i] } } \right)
\end{align*}
where $f(z) = z \Phi(z) + \phi(z)$ with $\phi$ and $\Phi$ being the
standard normal probability density and cumulative distribution
functions, respectively.  \citeN{Ryzhov2016} shows that EI does not
precisely achieve the OCBA allocation as the allocations to inferior
systems converge to zero. Under a Bayesian framework,
\citeN{Peng2017} propose a myopic allocation policy, called AOMAP, as
a new variant of EI, and show that AOMAP does achieve the OCBA allocation
when the variances are known.  Under this policy 
Step~3 becomes
\begin{align*} \label{sampleAOMAP}
\sample &= \argmax_{\x[i]  \in \X} \{\E{\max\{\m[i] - \At[i] ,0\} | \salg}\} 
 = \argmax_{\x[i]  \in \X} \left\{ \sqrt{  1 / \Pp[i] } \  
f\left(\frac{ \Pm[i] - \At[i]}{\sqrt{ 1 / \Pp[i] } } \right) \right \} 
\end{align*} 
where $ \At[i] = \Pm[\xStar]\ \ind[{\Pm[i] \neq \Pm[\xStar]}] +
(\Pm[\xStar] + \xit[\xStar] \Isa[\xStar]) \ \ind[{\Pm[i] =
\Pm[\xStar]}]$, and
\[\xit[\xStar] = \left( \sum_{\x[i]  \in \X \backslash
\{\x[\xStar]\}} \frac{\Iva[\xStar]\Iva[i]}{[\Pm[i] - \Pm[\xStar]]^4}
\right)^{-1/4}.  
\]
Notice that if $\xit[\xStar] = 0$, then $\At[i] =\Pm[\xStar]$, and
thus the expectation becomes $\ei[{\x[i]}]$.  Since EI is too
greedy in allocating to the best system, the additional term
adjusts the allocation to the best system to make it less
favorable as the number of iterations approaches infinity.  This
adjustment enables AOMAP to achieve the OCBA limiting allocation.  


\subsection{mCEI Policy}

Since EI does not fully capture the uncertainty in the output from a
stochastic simulation, \shortciteN{Salemi2019} introduced CEI in a
Gaussian Markov random field framework for discrete simulation
optimization.  For our R\&S problem, CEI for system $\x[i] \neq
\x[\xStar]$ at iteration $t$ is
\begin{align*}
\cei[{\x[i]}] & = \E{\max\{ \m[i] -  \m[\xStar],0\} | \salg}
=\sqrt{  1 / \Pp[i] + 1 / \Pp[\xStar] } \  f\left(\frac{ \Pm[i] - \Pm[\xStar]}{\sqrt{ 1 / \Pp[i] + 1 / \Pp[\xStar] } } \right).
\end{align*}
\citeN{Chen2019} present the mCEI policy for R\&S, which is
a modified version of the original CEI policy of
\shortciteN{Salemi2019}.  Under mCEI, $\sample = \x[\xStar]$ if 
\begin{equation}
\label{eq:cei.balance}
\left( \frac{\tnr[\xStar]}{\Isa[\xStar]} \right)^2 < 
\sum_{ \x[i]  \in \X \backslash \{\x[\xStar]\}
} \left( \frac{\tnr[i]}{\Isa[i]} \right)^2.
\end{equation} 
Otherwise, $\sample = \x[\xC]$ where $ \x[\xC] = \argmax_{\x[i] \neq
\x[\xStar]} \cei[{\x[i]}]$.  Condition~(\ref{eq:cei.balance}) enforces,
in the limit, the balance condition in the optimal allocation of
\citeN{Glynn2004}.


\subsection{TTTS Policy}

As TTTS involves more than a simple substitution for Step~3 in
Algorithm~\ref{alg:adaptiveAllocation}, we provide the new Step~3
as Algorithm~\ref{algo:TTTS}. The asymptotically best performance of
TTTS is obtained by tuning $\beta$ toward an optimal value. However,
\citeN{Russo2020} obtained good empirical performance by the simple
choice of $\beta = 1/2$.

\begin{algorithm}[h!]
  \caption{TTTS Step~3} \label{algo:TTTS}
  \begin{algorithmic}[]
  \State Sample $\widehat{\mu}_i \sim \normal(\Pm[i], 1/ \Pp[i])$ for $\x[i]\in\X$ and set $ I \gets \argmax_{i \in \X} \widehat{\mu}_i$. \Comment{Thompson sampling}
  \State Sample $B \sim \text{Bernoulli} (\beta)$.
  \If{$B=1$}
  \State $\sample = I$.
  \Else 
  \Repeat 
  \State Sample $\widehat{\mu} _j\sim \normal(\Pm[j], 1/ \Pp[j])$ for $\x[j]\in\X$ and set $J \gets \argmax_{j \in \X} \widehat{\mu}_j$.  
  \Until{$J\neq I$}
  \State $\sample = J$. 
\EndIf 
  \end{algorithmic}
\end{algorithm}

\section{gCEI POLICY}
\label{gCEI Algorithm and Main Results}

EI has been shown to be an effective search strategy in Bayesian
optimization of deterministic simulations; CEI extends EI to
stochastic simulation; and mCEI tailors CEI to obtain optimal
asymptotic performance in R\&S by insuring that the necessary balance
between simulating the best system and the rest is achieved in the
limit; a pure CEI policy never simulates the current sample best in
the next iteration.

One feature of CEI-based simulation-optimization methods such as GMIA
in \shortciteN{Salemi2019} is that CEI identifies promising solutions,
but not how many replications to expend on them.  gCEI grew out of an
ongoing investigation of employing CEI for that purpose by exploiting
its gradient with respect to the number of replications \tcb{treating
the number of replications as if it was continuous}. Here we use
it simply to decide how to allocate the next \emph{single}
replication, as with the other policies.

To derive an expression for the gradient of CEI, first notice that the
derivative of $f$ with respect to $z$ is $f^\prime(z) = \Phi(z)$.
Then, for $\x[i] \neq \x[\xStar]$,
$$\frac{\partial }{\partial \tnr[i]} \left( \frac{1}{\Pp[i] } \right)= -  \frac{\Iva[i] } {(\tnr[i] )^2 } 
\text{ and }\frac{ \partial \Pm[i] } {\partial  \tnr[i]} =\frac{ \partial \Em[i] } {\partial  \tnr[i]} = 0.$$ 
To simplify notation, let $\nu_i =  1 / \Pp[i] +  1 / \Pp[\xStar]$.
Then we have
\begin{align*}
\frac{\partial }{\partial \tnr[i]}\left(\frac{ \Pm[i] - \Pm[\xStar]}{\sqrt{ \nu_i} } \right) 
& = \frac{\left( \Em[i] - \Em[\xStar] \right) }{2 \nu_i\sqrt{ \nu_i} } \frac{\Iva[i] } {(\tnr[i] )^2 }. 
\end{align*}
Thus, the derivative of $\cei[{\x[i]}]$ with respect to $\tnr[i]$ is
\begin{eqnarray*}
\lefteqn{\frac{\partial \cei[{\x[i]}] } {\partial \tnr[i]}} \\ 
&=&- \frac{\Iva[i] }{2\sqrt{ \nu_i }(\tnr[i] )^2  } \ f\left(\frac{ \Em[i] - \Em[\xStar]}{\sqrt{\nu_i} } \right)
+ \sqrt{\nu_i} \left[ \frac{\left( \Em[i] - \Em[\xStar] \right) }{2
\nu_i\sqrt{ \nu_i} } \frac{\Iva[i] } {(\tnr[i] )^2 }  \right] \
f^\prime\left(\frac{ \Em[i] - \Em[\xStar]}{\sqrt{\nu_i} } \right)
\\[6pt]
&=& - \frac{\Iva[i] }{2\sqrt{ \nu_i }(\tnr[i] )^2  } \left[ \frac{
\Em[i] - \Em[\xStar]}{\sqrt{\nu_i} } \ \Phi\left(\frac{ \Em[i] -
\Em[\xStar]}{\sqrt{\nu_i} } \right) + \phi\left(\frac{ \Em[i] -
\Em[\xStar]}{\sqrt{\nu_i} } \right) \right. \\ 
&& \left. - \frac{\left( \Em[i] - \Em[\xStar] \right) }{ \sqrt{
\nu_i} } \ \Phi\left(\frac{ \Em[i] - \Em[\xStar]}{\sqrt{\nu_i} } \right) \right] \\[6pt]
&=&- \frac{\Iva[i] }{(\tnr[i] )^2  } \  \frac{1 }{2\sqrt{ \nu_i }}  \ \phi\left(\frac{ \Em[i] - \Em[\xStar]}{\sqrt{\nu_i} } \right)   \le 0 .
\end{eqnarray*}
Proceeding similarly, 
\begin{align*}
\frac{\partial \cei[{\x[i]}] } {\partial \tnr[\xStar]} 
& = - \frac{\Iva[\xStar] }{(\tnr[\xStar] )^2  } \  \frac{1 }{2\sqrt{ \nu_i }}  \ \phi\left(\frac{ \Em[i] - \Em[\xStar]}{\sqrt{\nu_i} } \right)   \le 0 
\end{align*}
whereas $\partial \cei[{\x[i]}] / \partial \tnr[j] =0 $ for $\x[j] \in
\X \backslash \{\x[i], \x[\xStar]\}$ because $\partial (\Pp[i])^{-1} /
\partial \tnr[j] =0$.

Since simulating the sample best $\x[\xStar]$ at iteration $t$ affects all
CEI's, the \emph{total} impact of simulating $\x[\xStar]$ is
$\sum_{\x[i] \in \X \backslash \{ \x[\xStar]\}} \partial \cei[{\x[i]}]
/ \partial \tnr[\xStar]$.  As lower CEI values are better, then from among the
systems other than $\x[\xStar]$, simulating
$$ \x[\xG] = \argmin_{\x[i]  \in \X \backslash \{ \x[\xStar]\}} \frac{\partial \cei[{\x[i]}] } {\partial \tnr[i]} $$
has potentially the most improvement.  To make a decision as to which
system to simulate next, $\x[\xStar]$ or $\x[\xG]$, we propose the
following condition:
\begin{equation*}
 \sum_{\x[i]  \in \X \backslash \{ \x[\xStar]\}} \frac{\partial
 \cei[{\x[i]}] } {\partial \tnr[\xStar]}  
\stackrel{?}{\le} \min_{\x[i]  \in \X \backslash \{ \x[\xStar]\}} \frac{\partial \cei[{\x[i]}] } {\partial \tnr[i]}  = \frac{\partial \cei[{\x[\xG]}] } {\partial \tnr[\xG]} .
\end{equation*}
If this condition holds, then the total impact of simulating
$\x[\xStar]$ is potentially greater than simulating $\x[\xG]$, and
thus we prefer simulating $\x[\xStar]$ to $\x[\xG]$, i.e., $\sample =
\x[\xStar]$.  On the other hand, if the condition does not hold, then
we prefer simulating $\x[\xG]$, i.e., $\sample =\x[\xG]$.  This leads
to the gCEI policy in Algorithm~\ref{gCEI}.

\begin{algorithm}[h!]
  \caption{gCEI Policy} \label{gCEI}
  \begin{algorithmic}[1]
  \State Let $\sample[0] = \x[i]$ for some $\x[i]\in\X$.
  Obtain $\obs[1]$ and update $\salg[1]$.
  Also, let $t \gets 1$.
  \While{$ t < \bdgt$} 
  \If{$$\sum_{\x[i]  \in \X \backslash \{\x[\xStar]\}} \partial
  \cei[{\x[i]}] / \partial \tnr[\xStar] \le   
  \min_{\x[i]  \in \X \backslash \{\x[\xStar]\}} \left\{  \partial
  \cei[{\x[i]}] / \partial \tnr[i] \right\}$$}
  \State $\sample = \x[\xStar]$.
  \Else 
  \State $\sample = \x[\xG]$ where $\x[\xG] = \argmin_{i \in \X
  \backslash \{\x[\xStar]\}}  \left\{  \partial \cei[{\x[i]}] /
  \partial \tnr[i] \right\}$. 
\EndIf
  \State  Obtain $\obs$ by simulating $\sample$, update $\salg[t+1] \gets \salg[t] \cup\{\sample, \obs\}$ and $t \gets t + 1$.
  \EndWhile
  \State Return $\x[\xStar][\bdgt] = \argmax_{i=1,\ldots, k}
  \{\Pm[i][\bdgt]\}$ as the selected best system.
  \end{algorithmic}
\end{algorithm}

Here we provide a sketch of the proof that gCEI converges to the
\citeN{Glynn2004} rate-optimal allocation.  First, it is easy to show
that as $R \rightarrow \infty$ the gCEI policy will simulate all
systems infinitely often.  Mimicing the analysis in
\citeN{Ryzhov2016}, we consider the deterministic behavior of gCEI
with the true means inserted for the estimates. This implies that
$k(t) = k$ for all $t$ and
\[
\frac{\partial \cei[{\x[i]}] } {\partial \tnr[i]} 
 = - \frac{\Iva[i] }{(\tnr[i] )^2  } \  \frac{1 }{2\sqrt{
\nu_i }}  \ \phi\left(\frac{ \m[i] - \m[k]}{\sqrt{\nu_i} }
\right) \mbox{  and  }
\frac{\partial \cei[{\x[i]}] } {\partial \tnr[k]} 
 = - \frac{\Iva[k] }{(\tnr[k] )^2  } \  \frac{1 }{2\sqrt{
\nu_i }}  \  \phi\left(\frac{ \m[i] - \m[k]}{\sqrt{\nu_i} } \right) .
\]
Consider the empirical allocation $\{\tnr[i][t]/t,\ i=1,2,\ldots,k\}$.
We know that it must have a convergent subsequence, $\tnr[i][t]/t
\stackrel{t\rightarrow \infty}{\longrightarrow} \alpha_i$ ; we show
that any such subsequence must converge to the rate-optimal
allocation \tcb{(a complete proof includes showing that the limit of
the subsequence is not $0$)}. Let $\nu_i^\prime = (\sigma_i^2/\alpha_i +
\sigma_k^2/\alpha_k) = \lim_{t \rightarrow \infty} t\nu_i$.

First consider the sub-subsequence on which the inequality in Step~3
holds. For such iterations 
\begin{equation}
\label{eq:cei.ratio}
\sum_{i \neq k} 
\frac{\sigma_k^2/(r_k(t))^2}{\sigma_j^2/(r_j(t))^2}
\sqrt{\frac{\nu_j}{\nu_i}} \times
\exp\left\{-\frac{1}{2}\left(
\frac{(\m[i] - \m[k])^2}{\nu_i} - \frac{(\m[j] -
\m[k])^2}{\nu_j}
\right)\right\} \ge 1 \mbox{ for any $j \neq k$.}
\end{equation}
However, as $t \rightarrow \infty$, the exponential term 
will go to $0$ or $\infty$ unless
\begin{equation}
\label{eq:GJ.diff}
\frac{(\m[i] - \m[k])^2}{\nu_i^\prime} = \frac{(\m[j] - \m[k])^2}{\nu_j^\prime} ,\ i \neq j \neq k.
\end{equation}
Thus, for Equation~(\ref{eq:cei.ratio}) to hold for any $j$,
Equation~(\ref{eq:GJ.diff}) must hold, which is the first of two
conditions for the rate-optimal allocation of \citeN{Glynn2004}.
Therefore, as $t$ increases, Inequality~(\ref{eq:cei.ratio})
becomes (after some manipulation)
\begin{equation}
\label{eq:simple.cei.ratio}
\sum_{i \neq k} 
\frac{\sigma_k^2/(r_k(t))^2}{\sigma_j^2/(r_j(t))^2}
\sqrt{\frac{1}{\nu_i}} \ge \sqrt{\frac{1}{\nu_j}} \mbox{ for any $j \neq k$.}
\end{equation}
Summing both sides over $j = 1,2,\ldots,k-1$, dividing out the common term, and
letting $t \rightarrow \infty$ gives
\begin{equation}
\label{eq:GJ.balance}
\sum_{\x[j] \neq k}
\left(\frac{\sigma_k}{\alpha_k}\right)^2 
\left(\frac{\alpha_j}{\sigma_j}\right)^2 \ge 1.
\end{equation}

Next consider the sub-subsequence on which inferior system $j \neq k$
is chosen in Step~3.  This reverses the inequality
in~(\ref{eq:simple.cei.ratio}), and must be true for each $j \neq k$.
Then a similar argument shows that the left-hand side
of~(\ref{eq:GJ.balance}) must be $\le 1$. Therefore, equality is
required, which is the second condition of \citeN{Glynn2004}.

\section{EMPIRICAL PERFORMANCE}
\label{Numerical Experiments}

We ran 16 experiments in total, including four different values of
number of systems $k \in \{5,10,20,30\}$.  For each $k$, we set $\m[i]
= c m_i$ where the $m_i$'s are prescaled true mean values provided in
Table~\ref{table:configurations} and $c$ is a scaling constant we
explain below.  In the slippage and ascending means configurations,
the systems have equal variances.  In the other two configurations the
means are ascending, but the variances are proportional to, and
inversely proportional to, the prescaled mean values.

\begin{table}[tb]
\newcommand*{\thead}[1]{\multicolumn{1}{c}{#1}}
\small
\centering
\caption{Configurations for experiments.}
{\def\arraystretch{1.3}\tabcolsep=15pt
\begin{tabular}{l l l}
\hline
\thead{Configuration} & \thead{Prescaled true mean values} & \thead{True standard deviations} \\
\hline
Slippage & $m_i = -1$ for $i \neq k$ and $m_k = 0$  & $\Isa[i] = 1$ \\  
Ascending mean & $m_i = \log (i)$  & $\Isa[i] = 1$ \\  
Ascending variance & $m_i = \log (i+1)$  & $\Isa[i] = \sqrt{m_i}$ \\
Descending variance & $m_i = \log (i+1)$  & $\Isa[i] = 1 / \sqrt{m_i}$ \\  \hline
\end{tabular}
}
\label{table:configurations}
\end{table}

In each experiment we first allocate 2 replications to each system
before applying any policy.  To create sensible cases, we scaled the
true means so that at least $r_0$ replications will be consumed before
the difference between the best and second-best systems is one
standard error of their estimated difference under the
\citeN{Glynn2004} rate-optimal policy.  Specifically, 
\[
\m[k] - \m[k-1] =  c (m_k - m_{k-1}) = \sqrt{
\frac{\Iva[k-1]}{r_0\alpha^\star_{k-1}} + \frac{\Iva[k]}{r_0\alpha^\star_{k}} }. 
\]
Thus, we control how quickly the best system becomes distinguishable from
the others.  To find $c$ satisfying the equation, we first calculate
the $\alpha^\star_i$ by solving the expression for the rate-optimal
allocation of \citeN{Glynn2004} with the $m_i$'s from
Table~\ref{table:configurations}. The constant $c$ does not change the
optimal allocation because scaling all $\m[i]$'s or all $\Isa[i]$'s
does not have any impact.  At the same time, we
want our total simulation budget $\bdgt$ to be large enough so that we
can  observe the convergence behavior of the policies.  We set
$r_0 = 20 k$ and $\bdgt = 100k$.  Lastly, we set the number of
macro-replications $M$ to 5000 to be able to estimate PICS to two
decimal places over a range of values. To measure the performance of
each policy, we report $\widehat{\text{PICS}}(t) = \sum_{\tau=1}^t
\ind[ {\xStar[\tau] \neq k}] / t$, $\widehat{\alpha}_k(t) = \tnr[k] / t$,
and the mean and standard deviation of $\m[k] - \m[\xStar]$, the
optimality gap, at each iteration $t$.

\begin{figure}[tb]
\centering
\small
\hspace{0.05cm}
\subfloat[PICS]{
\begin{tikzpicture}
\begin{axis}[axisStyle,  ylabel={$\widehat{\text{PICS}}(t)$}, xlabel={$t$},] 
\addplot graphics[xmin=0, xmax=1, ymin=0,ymax =1] {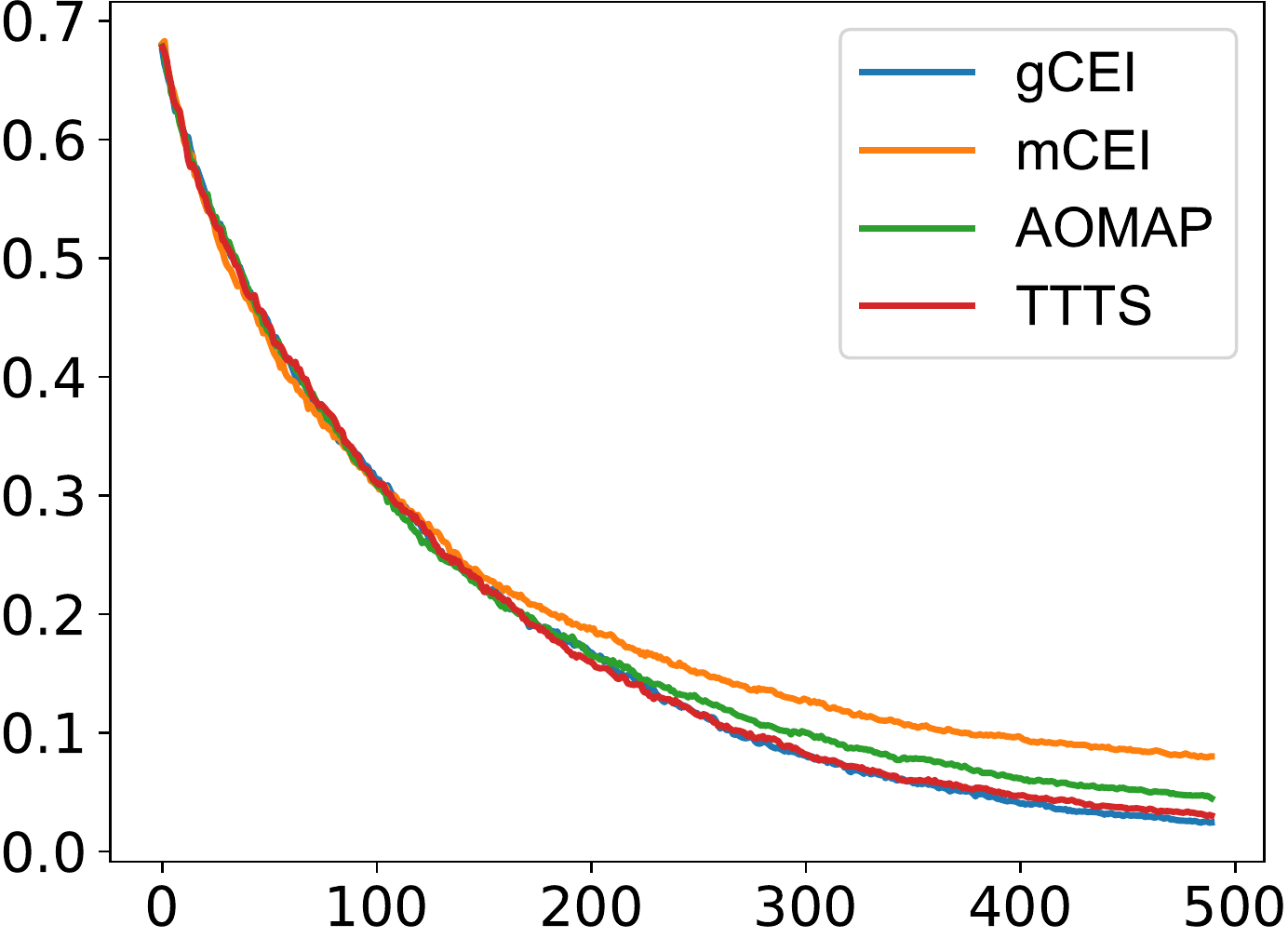}  ;
\end{axis}
\end{tikzpicture} }
\subfloat[Allocation to the best]{
\begin{tikzpicture}
\begin{axis}[axisStyle, ylabel={$\widehat{\alpha}_k(t)$}, xlabel={$t$}] 
\addplot graphics[xmin=0, xmax=1, ymin=0,ymax =1] {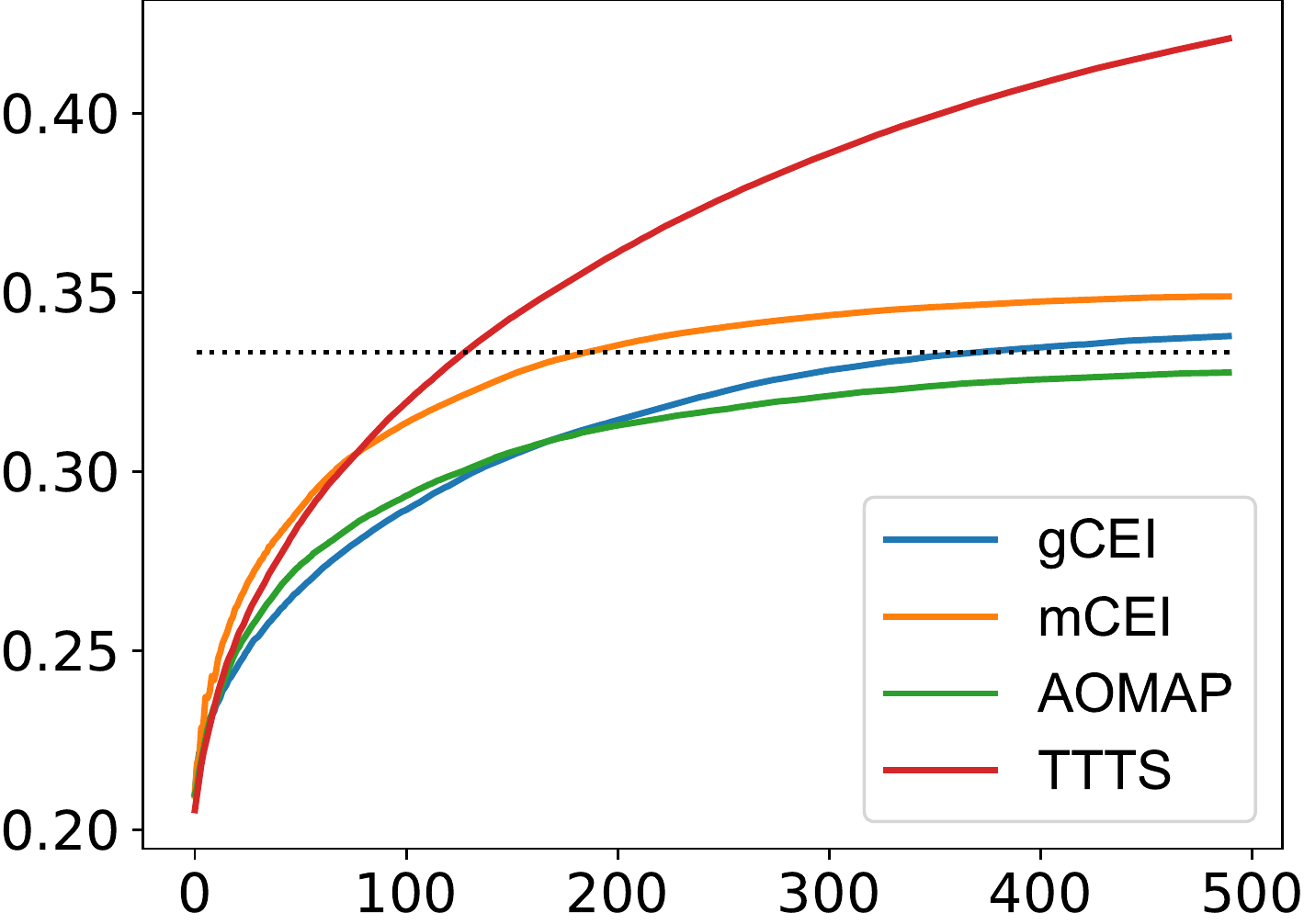}  ;
\end{axis}
\end{tikzpicture} } \\
\subfloat[Average gap]{
\begin{tikzpicture}
\begin{axis}[axisStyle, ylabel={Mean of $\m[k] - \m[\xStar]$}, xlabel={$t$}] 
\addplot graphics[xmin=0, xmax=1, ymin=0,ymax =1] {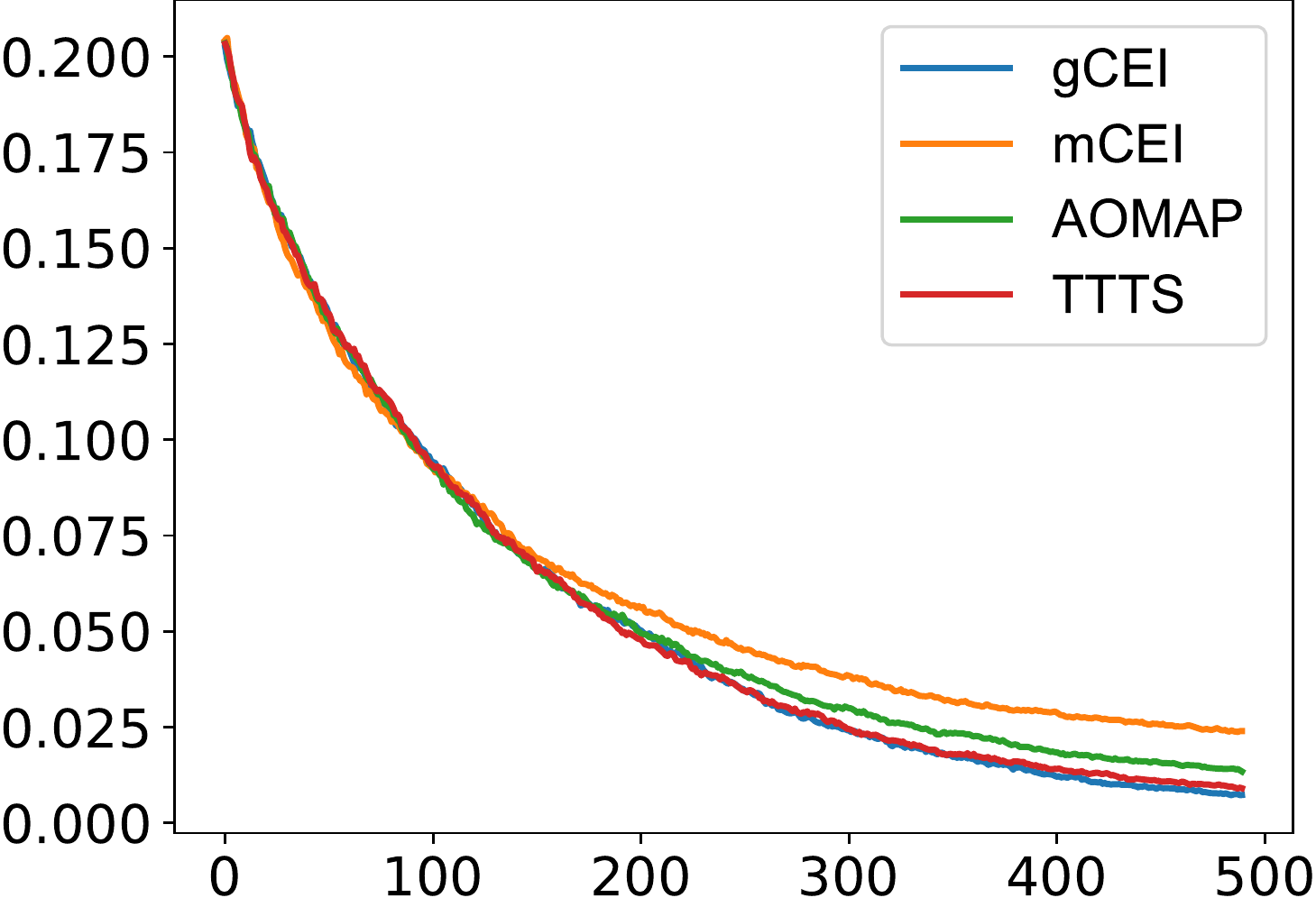}  ;
\end{axis}
\end{tikzpicture} } 
\subfloat[Std dev gap]{
\begin{tikzpicture}
\begin{axis}[axisStyle, ylabel={Std dev of $\m[k] - \m[\xStar]$}, xlabel={$t$}] 
\addplot graphics[xmin=0, xmax=1, ymin=0,ymax =1] {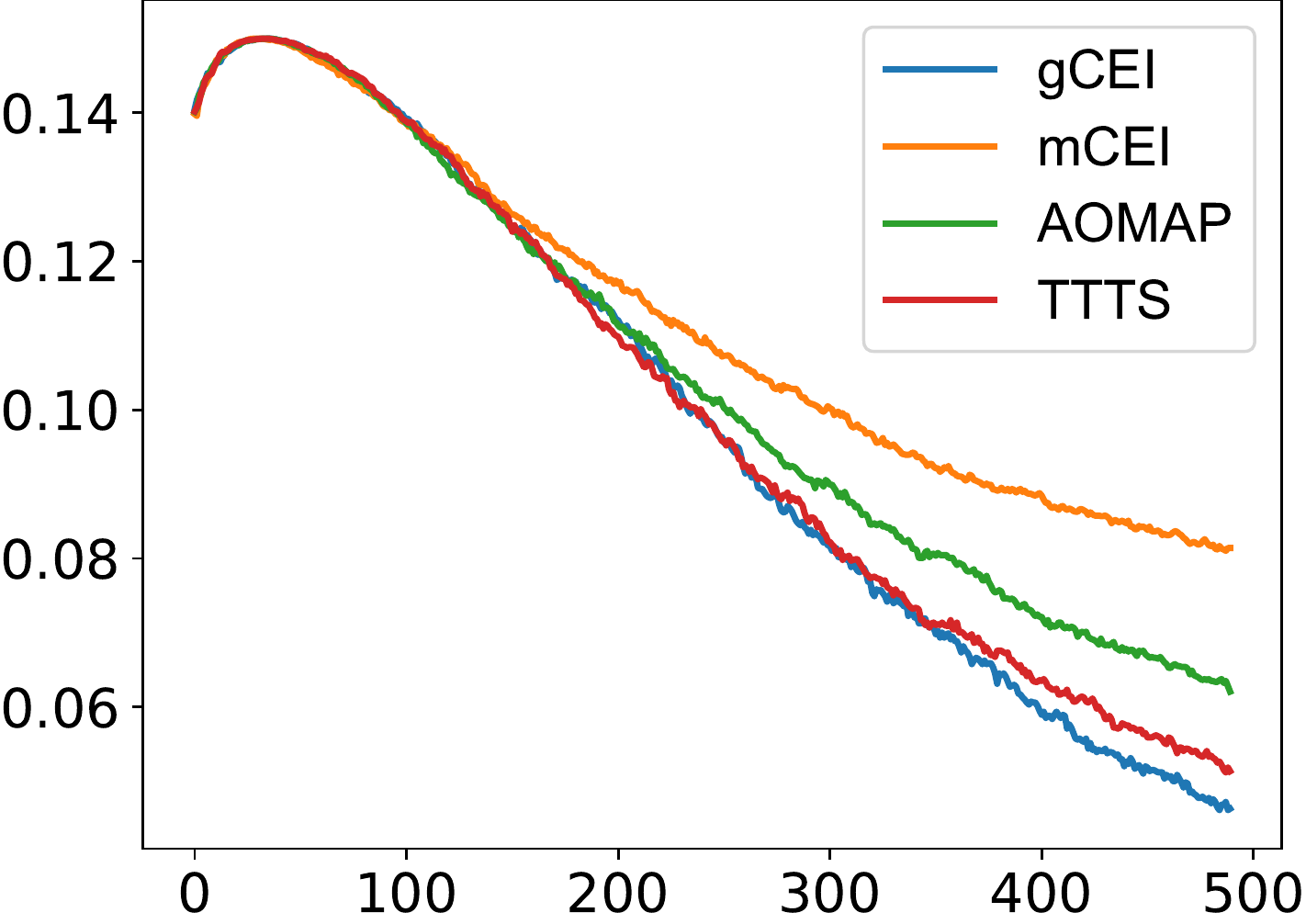}  ;
\end{axis}
\end{tikzpicture}}
\caption{The slippage configuration with $k=5$, $\bdgt=500$ and
$M=5000$. The dotted line in (b) is the Glynn \& Juneja optimal allocation
to the best system. }
\label{figure:slippage5}
\end{figure}

\begin{figure}[tb]
\centering
\small
\hspace{0.05cm}
\subfloat[PICS]{
\begin{tikzpicture}
\begin{axis}[axisStyle,  ylabel={$\widehat{\text{PICS}}(t)$}, xlabel={$t$},] 
\addplot graphics[xmin=0, xmax=1, ymin=0,ymax =1] {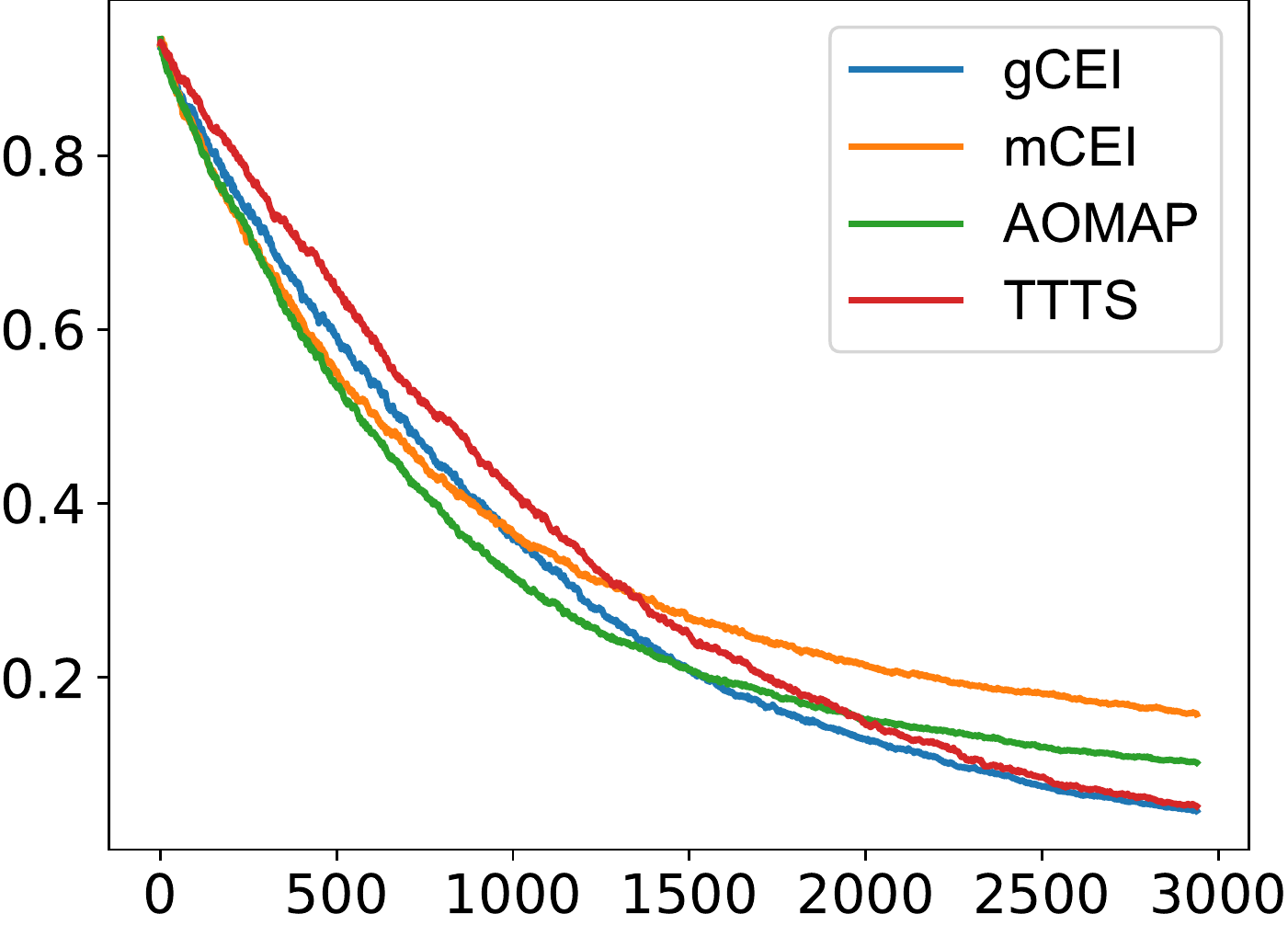}  ;
\end{axis}
\end{tikzpicture} }
\subfloat[Allocation to the best]{
\begin{tikzpicture}
\begin{axis}[axisStyle, ylabel={$\widehat{\alpha}_k(t)$}, xlabel={$t$}] 
\addplot graphics[xmin=0, xmax=1, ymin=0,ymax =1] {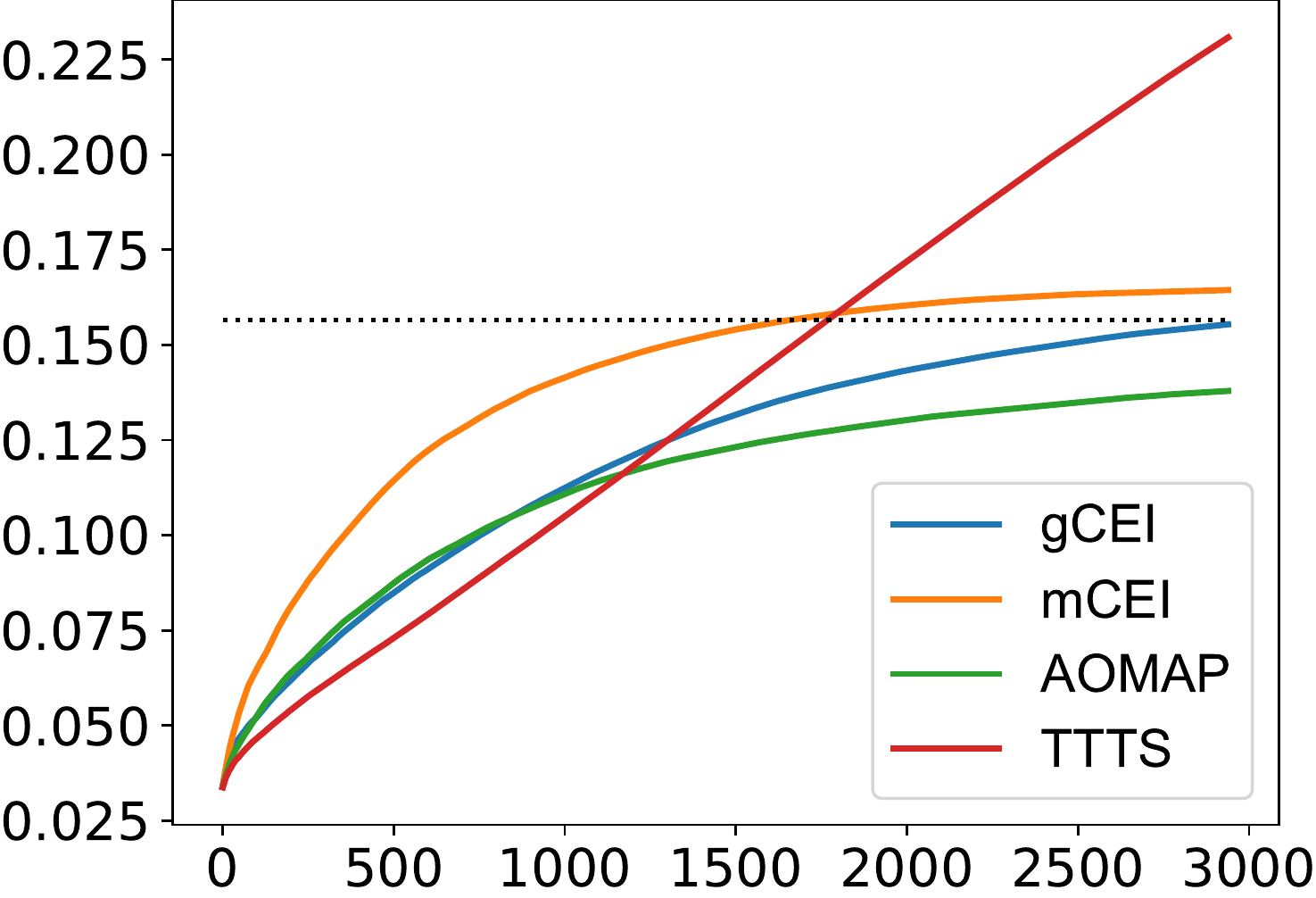}  ;
\end{axis}
\end{tikzpicture} } \\
\subfloat[Average gap]{
\begin{tikzpicture}
\begin{axis}[axisStyle, ylabel={Mean of $\m[k] - \m[\xStar]$}, xlabel={$t$}] 
\addplot graphics[xmin=0, xmax=1, ymin=0,ymax =1] {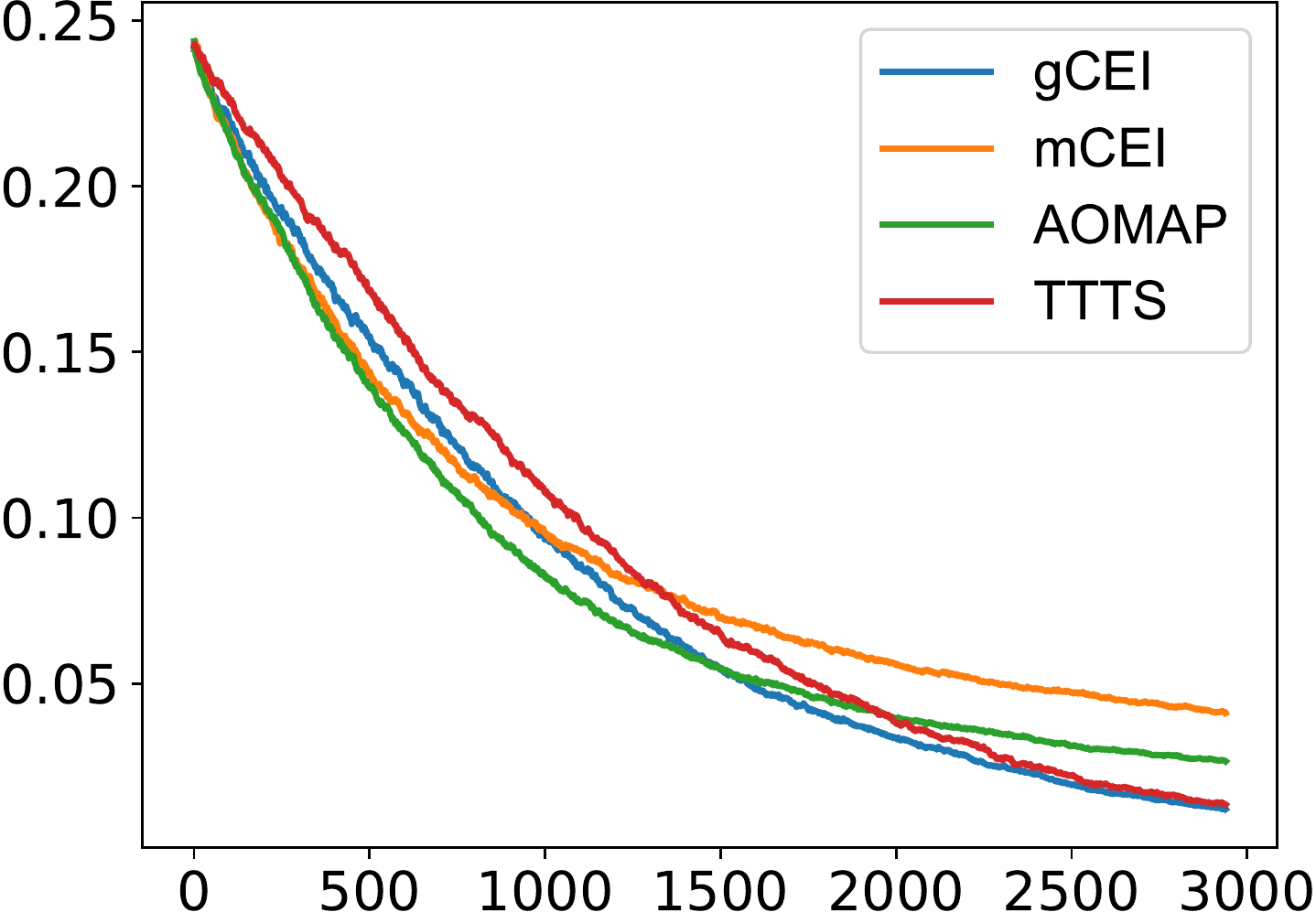}  ;
\end{axis}
\end{tikzpicture} } 
\subfloat[Std dev gap]{
\begin{tikzpicture}
\begin{axis}[axisStyle, ylabel={Std dev of $\m[k] - \m[\xStar]$}, xlabel={$t$}] 
\addplot graphics[xmin=0, xmax=1, ymin=0,ymax =1] {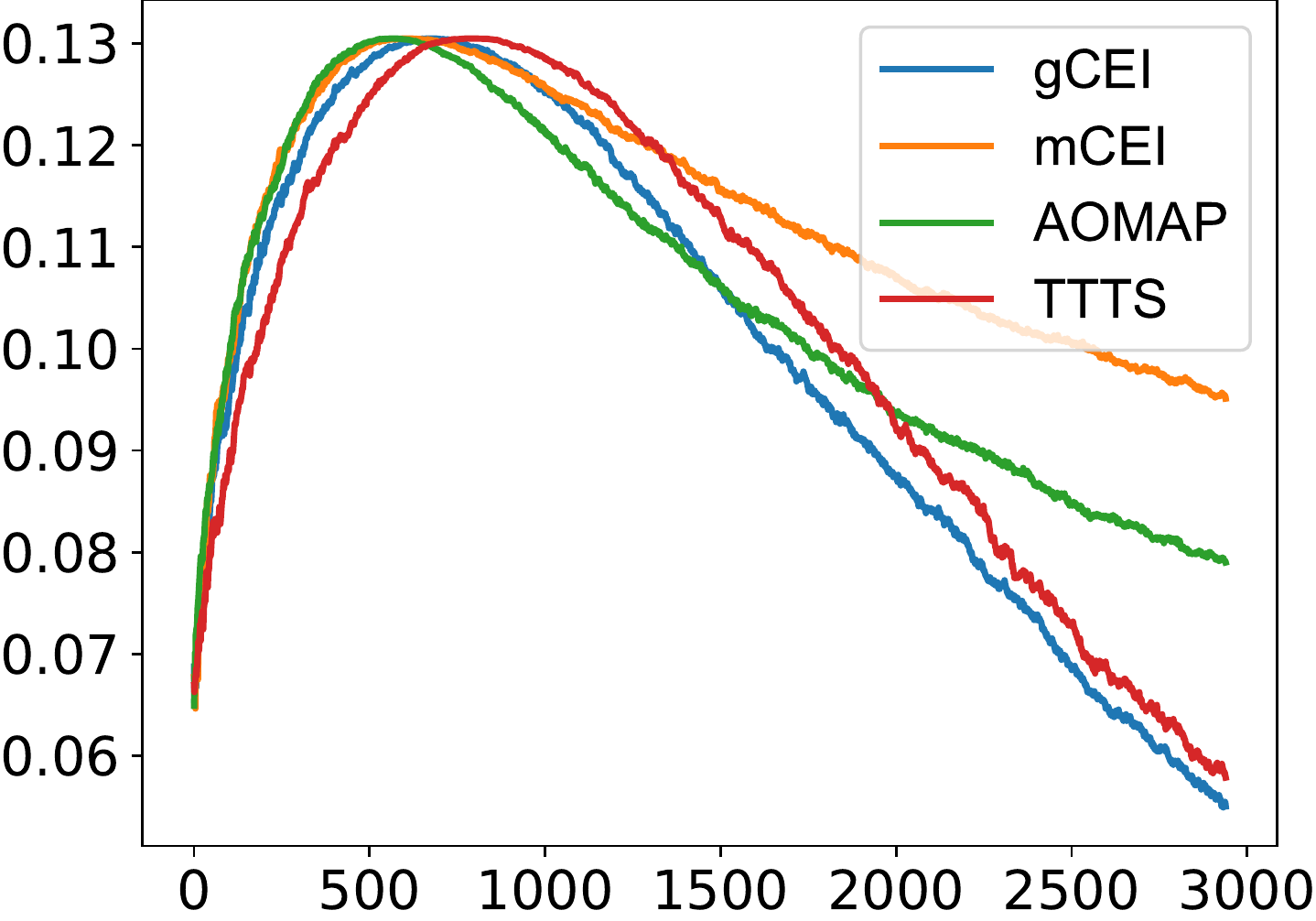}  ;
\end{axis}
\end{tikzpicture}}
\caption{The slippage configuration with $k=30$, $\bdgt=3000$ and
$M=5000$. The dotted line in (b) is the Glynn \& Juneja optimal allocation
to the best system.}
\label{figure:slippage30}
\end{figure} 

\begin{figure}[t]
\centering
\small
\hspace{0.05cm}
\subfloat[PICS]{
\begin{tikzpicture}
\begin{axis}[axisStyle,  ylabel={$\widehat{\text{PICS}}(t)$}, xlabel={$t$},] 
\addplot graphics[xmin=0, xmax=1, ymin=0,ymax =1] {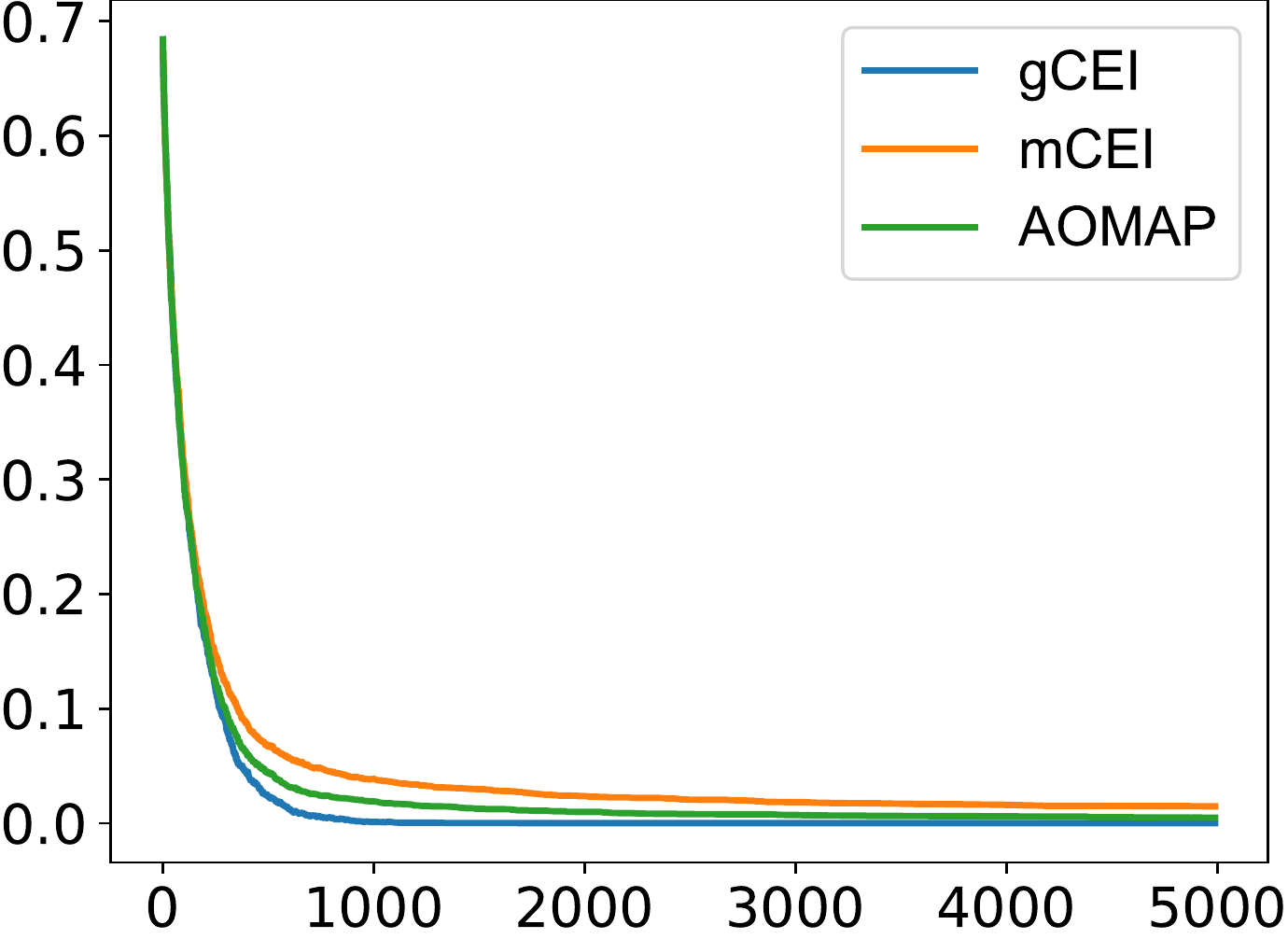}  ;
\end{axis}
\end{tikzpicture} }
\subfloat[Allocation to the best]{
\begin{tikzpicture}
\begin{axis}[axisStyle, ylabel={$\widehat{\alpha}_k(t)$}, xlabel={$t$}] 
\addplot graphics[xmin=0, xmax=1, ymin=0,ymax =1] {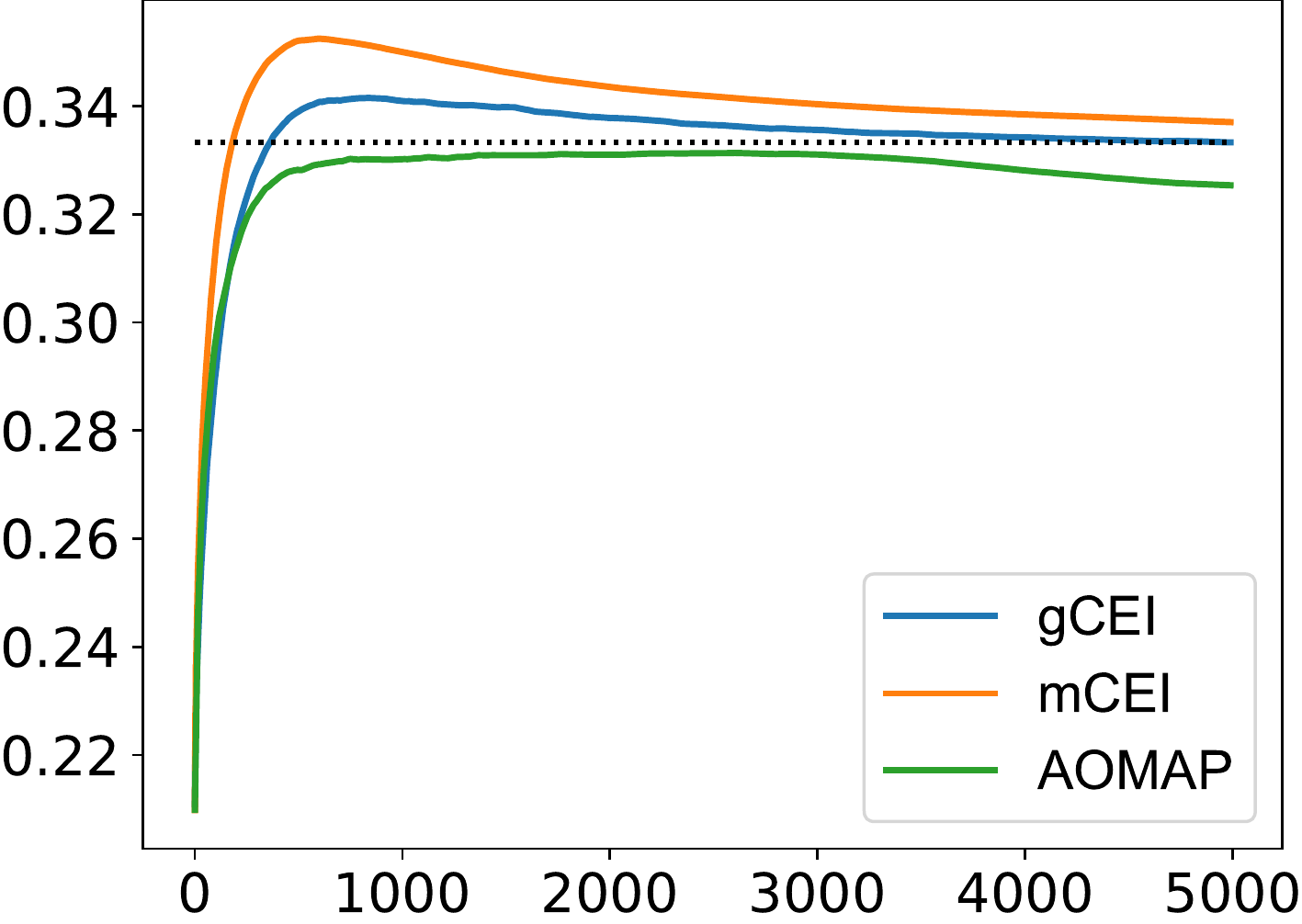}  ;
\end{axis}
\end{tikzpicture} } 
\caption{The slippage configuration with $k=5$, $\bdgt=5000$ and
$M=5000$. The dotted line in (b) is the Glynn \& Juneja optimal allocation
to the best system.}
\label{figure:slippage5long}
\end{figure}

\begin{figure}[h!]
\centering
\small
\hspace{0.05cm}
\subfloat[PICS]{
\begin{tikzpicture}
\begin{axis}[axisStyle,  ylabel={$\widehat{\text{PICS}}(t)$}, xlabel={$t$},] 
\addplot graphics[xmin=0, xmax=1, ymin=0,ymax =1] {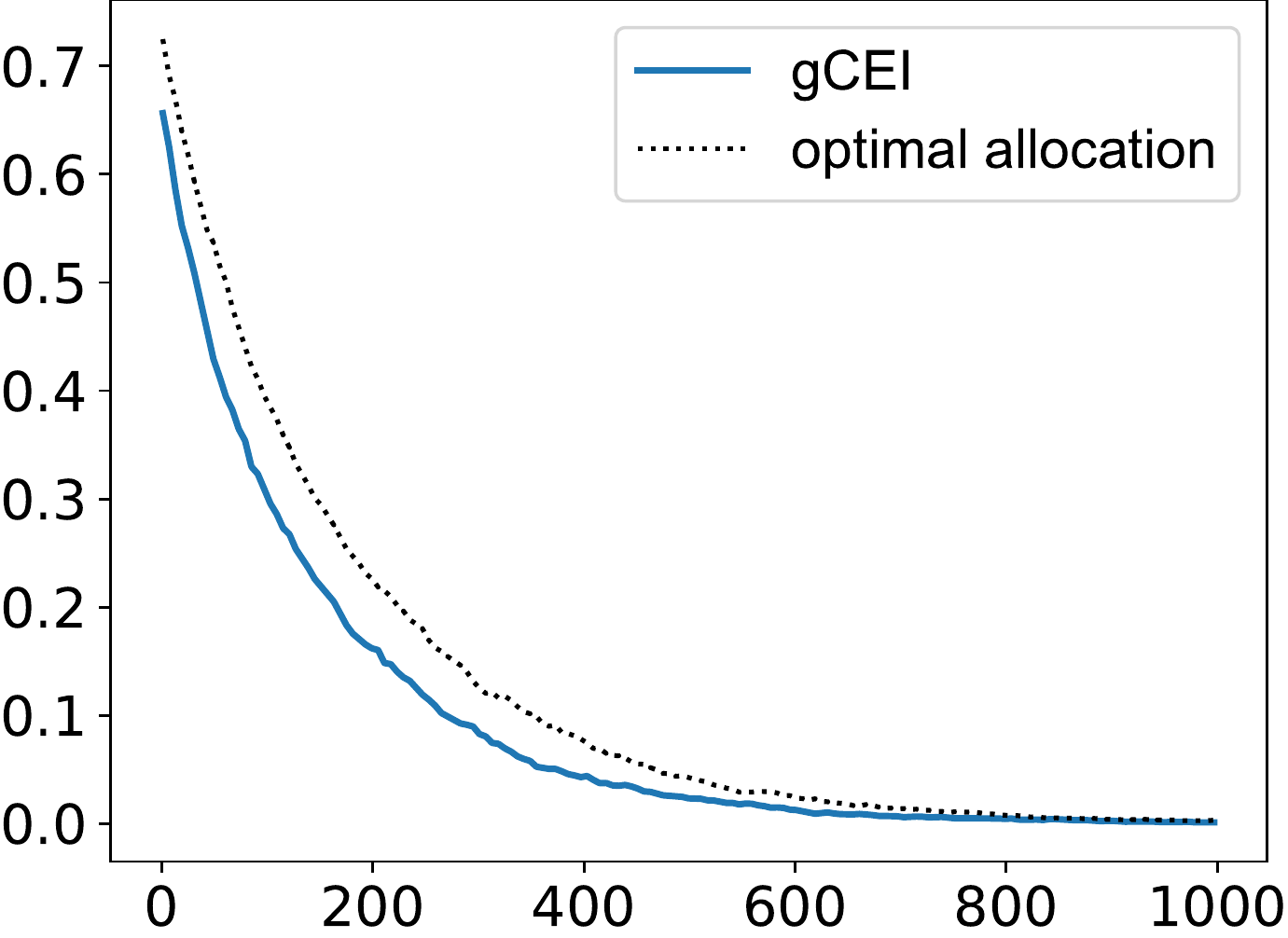}  ;
\end{axis}
\end{tikzpicture} }
\subfloat[Allocation to the best]{
\begin{tikzpicture}
\begin{axis}[axisStyle, ylabel={$\widehat{\alpha}_k(t)$}, xlabel={$t$}] 
\addplot graphics[xmin=0, xmax=1, ymin=0,ymax =1] {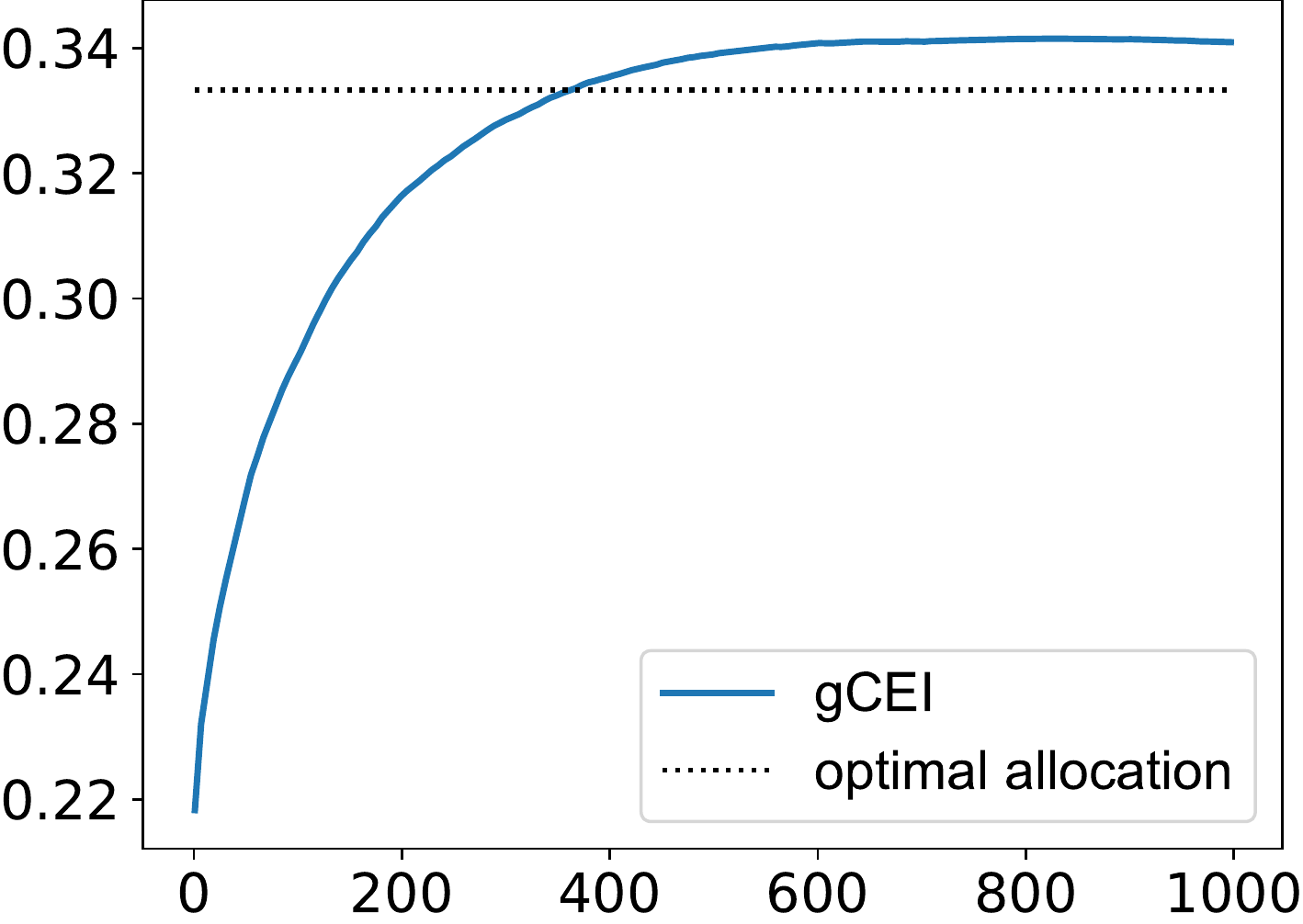}  ;
\end{axis}
\end{tikzpicture} } 
\caption{The slippage configuration with $k=5$, $\bdgt=1000$ and
$M=5000$. The dotted line in (b) is the Glynn \& Juneja optimal allocation
to the best system.}
\label{figure:slippage5opt}
\end{figure}

Figures~\ref{figure:slippage5}--\ref{figure:slippage30} exhibit
results for the slippage configurations with $k=5$ and $k=30$,
respectively.  To observe the tail behavior more closely,
Figure~\ref{figure:slippage5long} shows the results with a larger
budget of $\bdgt=5000$ for $k=5$.  We summarize our key observations:
Under each policy, PICS converges to zero as expected.  However, the
convergence behaviors are not always the same, and it appears that
gCEI performs as well or better than the other policies.
Both gCEI and mCEI first overshoot the asymptotically optimal
allocation of \citeN{Glynn2004} for the best system, but then
converge as expected in the long run.  gCEI tends to
allocate less to the best system than mCEI.  On the other hand, TTTS
allocates much more to the best system, which makes sense given its
heritage in MAB and minimizing regret.  Remember that the limiting
allocations for AOMAP and TTTS are \emph{not} those of Glynn and
Juneja, so we do not expect the same allocations. 

For the slippage configuration the mean optimality gap is a scaled
version of PICS because the gap is the same whenever any inferior
system is selected as the best. The standard deviation of the
optimality gap shows what might be considered unexpected behavior as
it first increases and then decreases.  This is because the best
system is not distinguishable with a small budget and the inferior
systems all have the same mean values.  As the best system becomes
more recognizable, the variability increases up to a point, then
decreases as each policy becomes more sure of the identity of the best
system. 


To understand how the dynamic policies behave relative to employing
the \emph{asymptotically} optimal allocation of \citeN{Glynn2004} from the
beginning, we compare gCEI with an unrealistic policy where the Glynn
and Juneja allocation is known and is applied starting from the first
iteration in the slippage configuration with $k=5$.  More
specifically, under this unrealistic policy, two replications should
be allocated to the best system for each  replication allocated to
an inferior system.  Figure~\ref{figure:slippage5opt} exhibits the
result of this comparison where we only report iterations that are a
multiple of six.  gCEI performs better than this unrealistic policy
even though it overallocates to the best system for a while. This
emphasizes that the rate-optimal allocations address large-sample, not
small-sample, behavior.

\begin{figure}[h!]
\centering
\small
\hspace{0.05cm}
\subfloat[PICS]{
\begin{tikzpicture}
\begin{axis}[axisStyle,  ylabel={$\widehat{\text{PICS}}(t)$}, xlabel={$t$},] 
\addplot graphics[xmin=0, xmax=1, ymin=0,ymax =1] {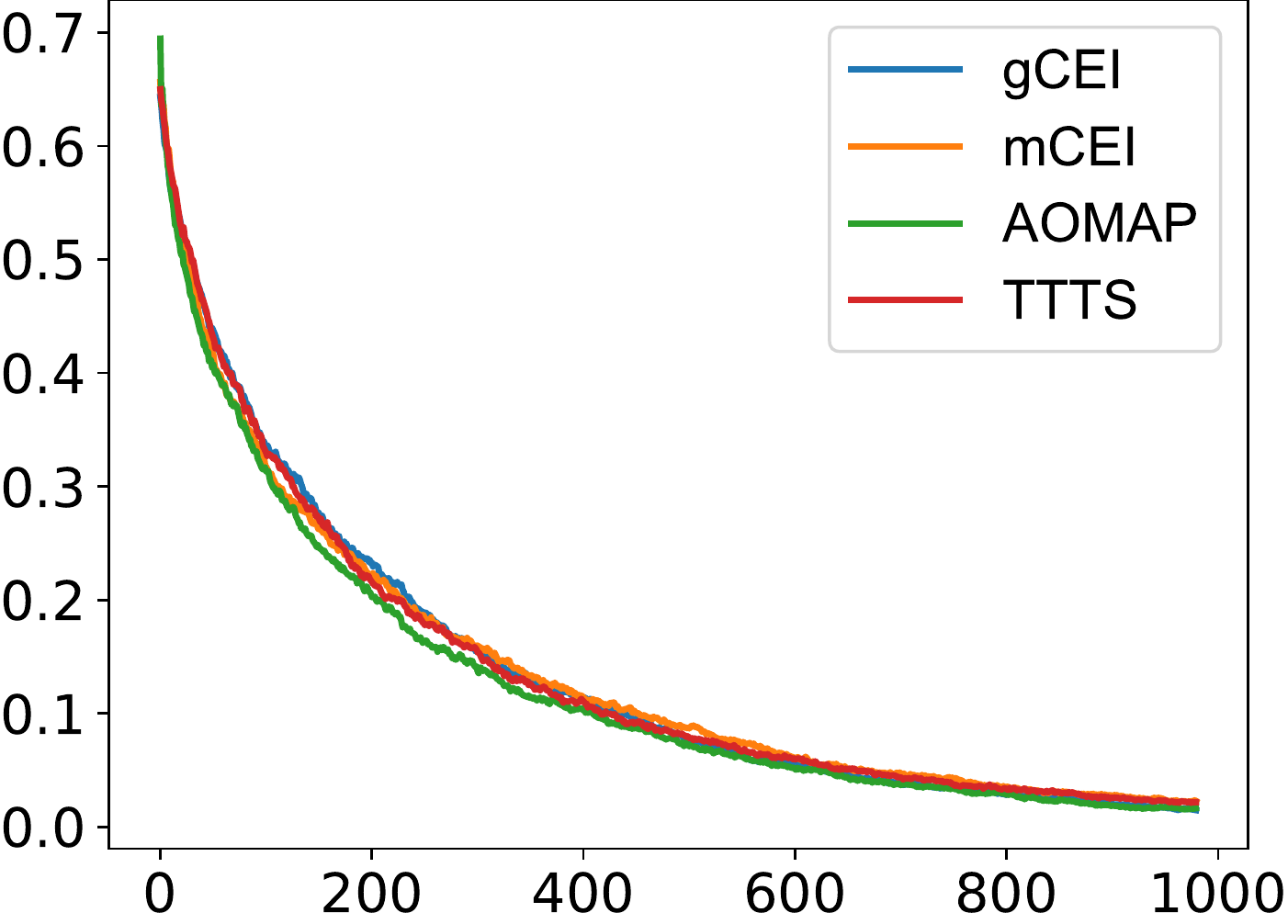}  ;
\end{axis}
\end{tikzpicture} }
\subfloat[Allocation to the best]{
\begin{tikzpicture}
\begin{axis}[axisStyle, ylabel={$\widehat{\alpha}_k(t)$}, xlabel={$t$}] 
\addplot graphics[xmin=0, xmax=1, ymin=0,ymax =1] {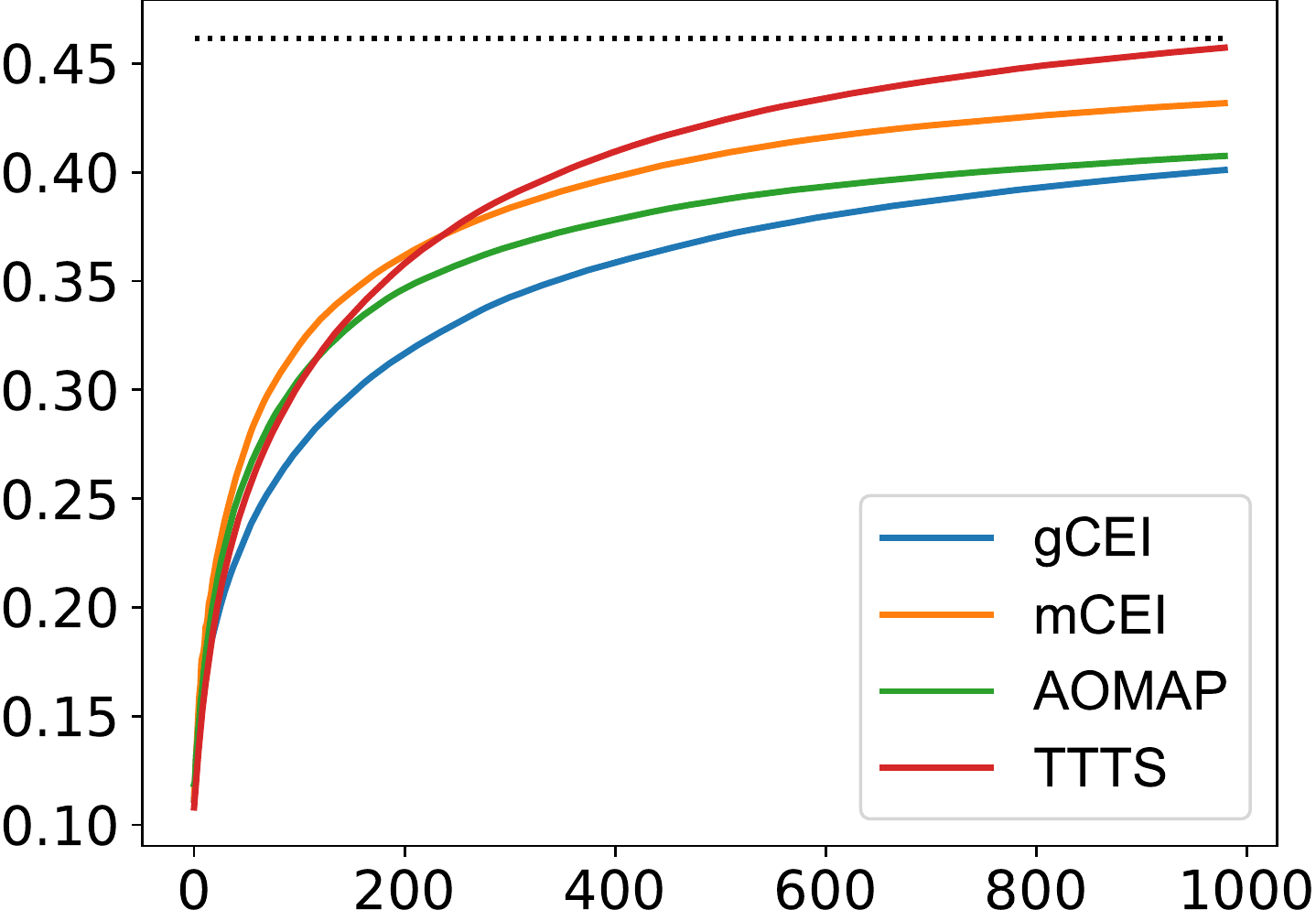}  ;
\end{axis}
\end{tikzpicture} } \\
\subfloat[Average gap]{
\begin{tikzpicture}
\begin{axis}[axisStyle, ylabel={Mean of $\m[k] - \m[\xStar]$}, xlabel={$t$}] 
\addplot graphics[xmin=0, xmax=1, ymin=0,ymax =1] {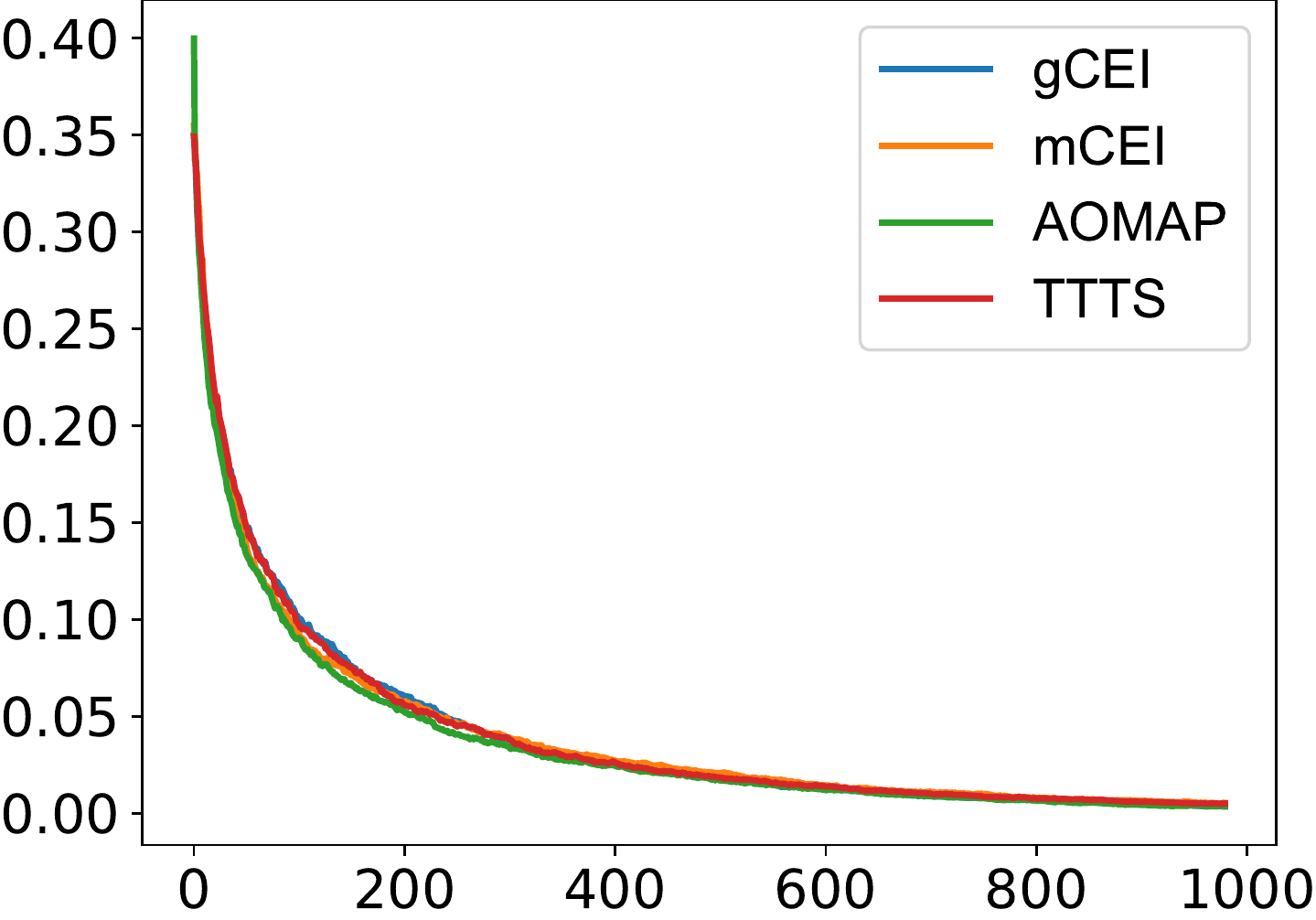}  ;
\end{axis}
\end{tikzpicture} } 
\subfloat[Std dev gap]{
\begin{tikzpicture}
\begin{axis}[axisStyle, ylabel={Std dev of $\m[k] - \m[\xStar]$}, xlabel={$t$}] 
\addplot graphics[xmin=0, xmax=1, ymin=0,ymax =1] {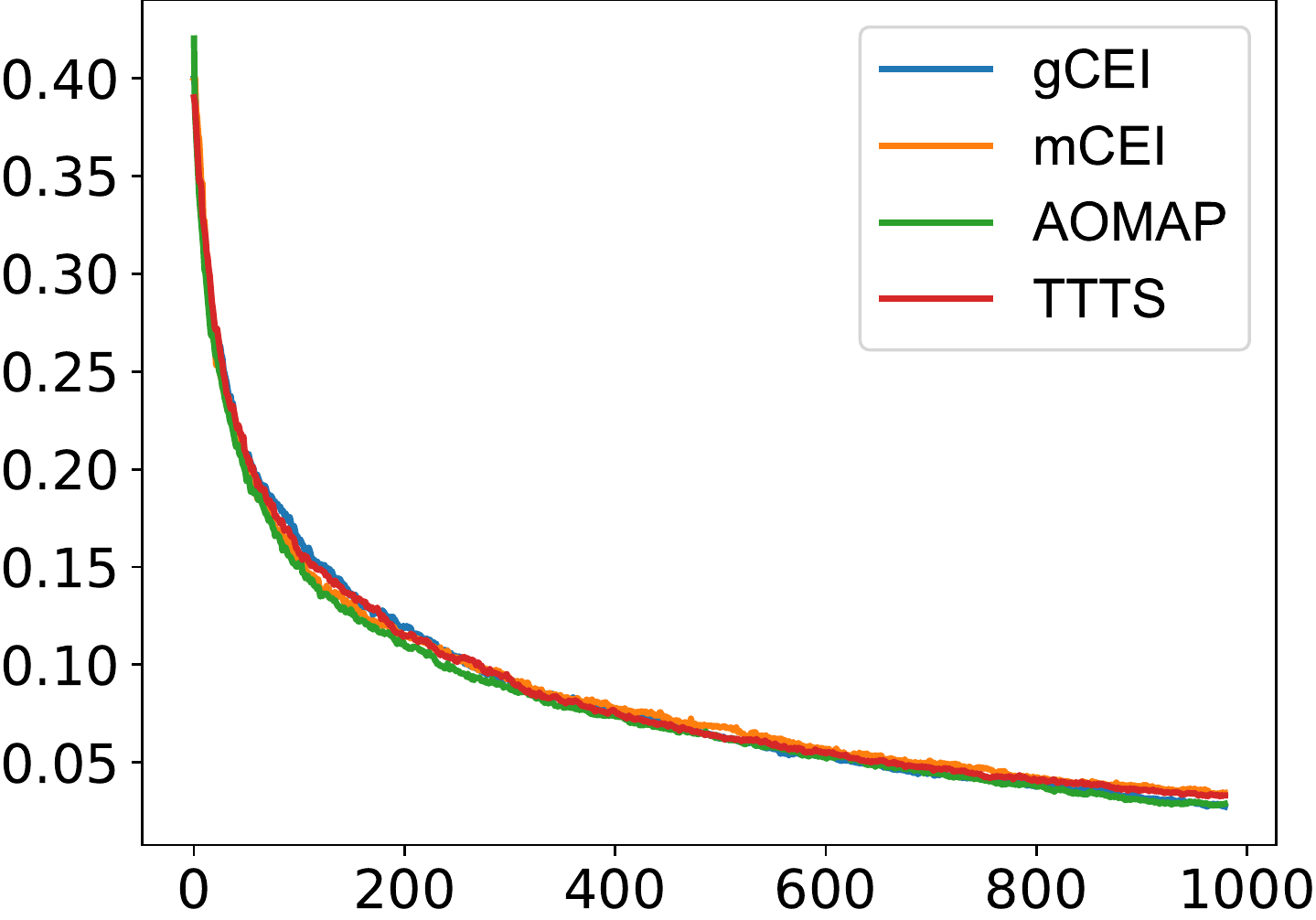}  ;
\end{axis}
\end{tikzpicture}} 
\caption{The ascending variance configuration with $k=10$,
$\bdgt=1000$ and $M=5000$. The dotted line in (b) is the Glynn \& Juneja optimal allocation
to the best system.}
\label{figure:ascendingVariance}
\end{figure}

Lastly, Figure~\ref{figure:ascendingVariance} exhibits results for the
ascending variance configuration with $k=10$.  We do not report the
results for the other configurations as they are so similar to this
one.  Here all policies perform similarly, based on our
metrics.  The only difference appears in their allocations to the best
system.  Similar to the slippage configuration, gCEI allocates less to
the best system than mCEI.  However, in contrast to the slippage
configuration, mCEI and gCEI do not overshoot the asymptotically
optimal allocation of \citeN{Glynn2004}.

The slippage configuration is certainly unrealistic, but it represents
a situation in which there are many close competitors to the best. In
this setting gCEI seems to have some advantages. When the means are
ascending it appears to be easier for all policies to control the PICS
and optimality gap because the inferior systems are easier to
identify; we found that all policies tended to allocate the majority
of their replications to the top two systems in these settings.

\section{CONCLUSIONS}
\label{Conclusion}

In this paper we examined three recent policies, and one new policy,
for assigning replications to systems in the fixed-budget R\&S
problem. All of the policies adapt as they obtain additional
simulation outputs, and each policy achieves a form of optimal
allocation as the budget increases; they differ in their definition of
``optimal'' and their small-sample behavior. Looking at PICS, and the
mean and standard deviation of the optimality gap at termination, gCEI
appears to perform as well or better than AOMAP, mCEI and TTTS.
\tcb{Our comparisons did not consider computational effort (other than
replications) or the ability to stop with a prespecified PCS, measures
that also distinguish R\&S procedures.}


\section*{ACKNOWLEDGEMENTS}

This research was partially supported by National Science Foundation
Grant Number DMS-1854562.

\footnotesize

\bibliographystyle{wsc}

\bibliography{References}

\section*{AUTHOR BIOGRAPHIES}

\noindent {\bf HARUN AVCI} is a PhD candidate in the Department of
Industrial Engineering and Management Sciences at Northwestern
University.  His research interests include simulation optimization
and simulation methodology.  His e-mail address is
\email{harun.avci@u.northwestern.edu}.  \\

\noindent {\bf BARRY L NELSON} is the Walter P.\ Murphy Professor in
the Department of Industrial Engineering \& Management Sciences at
Northwestern University.  He is a Fellow of INFORMS and
IIE, and is the
author of \textit{Foundations and Methods of Stochastic Simulation: A
First Course}, from Springer.  His e-mail and web addresses are
\email{nelsonb@northwestern.edu} and
\url{http://www.iems.northwestern.edu/~nelsonb}, respectively.  \\

\noindent {\bf ANDREAS W\"{A}CHTER} is a Professor in the Department
of Industrial Engineering and Management Sciences at Northwestern
University.  His research centers on the design, analysis,
implementation, and application of numerical algorithms for nonlinear
optimization.  He is a recipient of the J. H. Wilkinson Prize for
Numerical Software for the Ipopt open-source optimization package, and
a Fellow of SIAM.  His e-mail address is
\email{andreas.waechter@northwestern.edu}. 

\end{document}